\documentclass{article}
\usepackage{piner}
\usepackage[figuresright]{rotating}
\begin{document}
 
\title{Kinematics of the Parsec-Scale Relativistic Jet in Quasar 3C\,279: 1991 -- 1997}

\author{A. E. Wehrle\altaffilmark{1}, B. G. Piner\altaffilmark{1,2}, S. C. Unwin\altaffilmark{1},
A. C. Zook\altaffilmark{3}, W. Xu\altaffilmark{4}, A. P. Marscher\altaffilmark{5},
H. Ter\"asranta\altaffilmark{6}, \& E. Valtaoja\altaffilmark{7,8}}

\altaffiltext{1}{Jet Propulsion Laboratory,
California Institute of Technology, 4800 Oak Grove Drive, Pasadena, CA
91109; Ann.E.Wehrle@jpl.nasa.gov, B.G.Piner@jpl.nasa.gov; Stephen.C.Unwin@jpl.nasa.gov}

\altaffiltext{2}{Department of Physics and Astronomy, Whittier College,
13406 E. Philadelphia Street, Whittier, CA 90608; gpiner@whittier.edu} 

\altaffiltext{3}{Department of Physics and Astronomy, Pomona
College, Claremont, CA 91711; azook@pomona.edu}

\altaffiltext{4}{Infrared Processing and Analysis Center, Jet Propulsion Laboratory,
California Institute of Technology 100-22, Pasadena CA 91125}

\altaffiltext{5}{Institute for Astrophysical Research, Boston
University, 725 Commonwealth Avenue, Boston MA 02215; marscher@bu.edu}

\altaffiltext{6}{Mets\"ahovi Radio Observatory, Mets\"ahovintie 114,
02540 Kylm\"al\"a, Finland; hte@kurp.hut.fi}

\altaffiltext{7}{Tuorla Observatory, V\"ais\"al\"antie 20, FIN-21500 Piikki\"o, Finland}

\altaffiltext{8}{Department of Physics, Turku University, FIN-20100 Turku, Finland; valtaoja@deneb.astro.utu.fi}

\begin{abstract}
We present results of long-term high-frequency VLBI monitoring of the 
relativistic jet in 3C\,279, consisting of 18 epochs at 22 GHz from 1991 to 1997
and 10 epochs at 43 GHz from 1995 to 1997.  Three major results of this study are:
apparent speeds measured for six superluminal components range from
4.8 to 7.5$c$ ($H_{0}$=70 km s$^{-1}$ Mpc$^{-1}$, $q_{0}$=0.1),
variations in the total radio flux
are due primarily to changes in the VLBI core flux, and
the uniform-sphere brightness temperature of the VLBI core is $\sim1\times10^{13}$~K
at 22 GHz after 1995, one of the highest direct estimates of a brightness temperature.
If the variability brightness temperature measured for 3C\,279 by L\"{a}hteenm\"{a}ki \& Valtaoja
is an actual value and not a lower limit, then the rest-frame brightness temperature of 3C\,279 is
quite high and limited by inverse Compton effects rather than equipartition.

The parsec-scale morphology of 3C\,279 consists of a bright, compact VLBI core,
a jet component (C4) that moved from $\sim$2 mas to $\sim$3.5 mas from the core
during the course of our monitoring,
and an inner jet that extends from the core to a stationary component, C5, at $\sim$1 mas from the core.
Component C4 followed a curved path, and we reconstruct its three-dimensional trajectory
using polynomial fits to its position versus time.
Component C5 faded with time,
possibly due to a previous interaction with C4 similar to interactions seen in simulations by G\'{o}mez et al.
Components in the inner jet are relatively short-lived, and fade by the time they reach
$\sim$1 mas from the core.  The components have different speeds and 
position angles from each other, but these differences do not match the differences
predicted by the precession model of Abraham \& Carrara.  
Although VLBI components
were born about six months prior to each of the two observed $\gamma$-ray high states,
the sparseness of the $\gamma$-ray data prevents a statistical analysis of
possible correlations.
\end{abstract}


\section{INTRODUCTION}
\label{sec-intro}

The quasar 3C\,279 ($z = 0.536$) is one of the archetypal superluminal
radio sources (Cotton et al. 1979).  At $\gamma$-ray energies, the light curve
of 3C\,279 has been sampled intermittently since the launch of the
Compton Observatory in 1991; it is one of the brightest EGRET
quasars (Hartman et al. 1999). 3C\,279 is also well known as an optically violent variable
(OVV), with large and rapid outbursts (Webb et al. 1990).  Strong
variability on timescales shorter than one day is observed in
high-energy bands (Wehrle et al. 1998; Lawson, McHardy, \& Marscher 1999).

Correlations between the variability seen over the entire
electromagnetic spectrum have proved elusive.  Variability occurs on a
variety of timescales, especially at high energies, and time sampling
has been adequate to track the variations only at radio, millimeter, and
X-ray bands, and as far as practicable in optical bands.  There are
strong theoretical motivations for the search for correlations.  The
two-humped overall spectral energy distribution is most naturally
explained as a combination of synchrotron radiation for the radio
through optical-uv region, and inverse Compton emission at higher
energies (Maraschi et al. 1994; Wehrle 1999).  The synchrotron 
and inverse Compton emission is generally
thought to be associated with a jet of relativistic electrons; however,
the source of the seed photons for inverse Compton scattering is a
matter of considerable debate (e.g. Maraschi, Ghisellini, \& Celotti 1992; 
Sikora, Begelman, \& Rees 1994).
Distinguishing the differing mechanisms involves a full understanding
of the time correlations in the different energy bands.  A consensus
is emerging that for GeV-peaked blazars, the seed photons upscattered
to x-ray and $\gamma$-ray energies originate outside the jet (e.g., in the
accretion disk or broad line region clouds) with a minor
contribution from synchrotron photons (Urry 1999).  A
full understanding may have to await the next generation of 
$\gamma$-ray satellite observatories. 

In the radio regime, the variability timescale is longer, and flux
monitoring at 4.8, 8.4 and 14.5 GHz, complete with polarization data,
has been obtained at the University of Michigan Radio Observatory 
(e.g. Aller et al. 1985).  Monitoring at 22 and 37 GHz has been done at Mets\"{a}hovi
Observatory (Ter\"{a}sranta et al. 1992; Ter\"{a}sranta et al. 1998). Less frequent monitoring has been
performed at SEST (90 and 230 GHz) (Tornikoski et al. 1996)
and at the JCMT (230 GHz) (Marscher et al. 1999).

The time variability of 3C\,279's VLBI structure has been studied by
several groups, beginning with the earliest days of the VLBI technique
itself (Knight et al. 1971; Whitney et al. 1971; Cohen et al. 1971).  
Most of these observations were made using the ad-hoc
US and European VLBI Networks, with observations at intervals of about 1
year (Unwin et al. 1989 [hereafter U89]; Carrara et al. 1993 [hereafter C93]; 
Abraham \& Carrara 1998).  With the advent of the NRAO
Very Long Baseline Array\footnote{The National Radio Astronomy Observatory
is a facility of the National Science Foundation operated
under cooperative agreement by Associated Universities, Inc.} (VLBA), more frequent
monitoring began in 1991, with an emphasis on higher radio
frequencies (22 and 43 GHz).  This paper presents results from this more
frequent monitoring; preliminary results from this monitoring have been presented
by Wehrle, Unwin, \& Zook (1994), Wehrle et al. (1996), and Unwin et al. (1998).  
Polarization-sensitive VLBI images of 3C\,279 have also been made by
Lepp\"{a}nen, Zensus, \& Diamond (1995), Cawthorne \& Gabuzda (1996), Lister, Marscher, \& Gear (1998),
Lister \& Smith (2000), and Homan \& Wardle (1999) (who detect a significant
component of circular polarization).  Space VLBI observations at 5 and 1.6 GHz have
been performed with the VLBI Space Observatory Programme (VSOP) since 1998; first
results are reported by Piner et al. (2000a).

The highest angular resolution achieved on 3C\,279 is 50 $\mu$as, at
86 GHz (Rantakyr\"{o} et al. 1998), and VLBI fringes have been detected
up to frequencies of 215~GHz (Krichbaum et al. 1997).
In images from 1990 and 1992, Rantakyr\"{o} et al. (1998)
showed a narrow string of components within about 1 mas of the
core.  Rantakyr\"{o} et al. speculate that most of the ``missing'' flux lies in
a more extended jet which is resolved out by their 50 $\mu$as beam.

In this paper, we present the results of a long-term VLBI monitoring
campaign on 3C\,279.  The data comprise VLBI images at 22
GHz over the period 1991 - 1997 (a total of 18 epochs) and 
at 43 GHz over the period 1995 - 1997 (10 epochs). 
In Section~\ref{sec-vlbiobs} of this paper we present the VLBI observation series,
explaining how the data were collected, calibrated, and analyzed.
Section~\ref{sec-imaging} presents the VLBI images.
Section~\ref{sec-slmotion} discusses the superluminal motion visible in the image
sequence, and shows that different regions of the jet show
qualitatively different evolution.  Section~\ref{sec-radioevolution} analyzes the flux
density and spectral evolution of the radio core and components in the VLBI jet.  We present our
conclusions in Section~\ref{conclusions}.  In a subsequent paper (Piner et
al. 2000b) we will combine synchrotron self-Compton models with
our VLBI data and X-ray data to further constrain the jet kinematics.
Throughout the paper we assume $H_{0}$=70 km s$^{-1}$ Mpc$^{-1}$ and $q_{0}$=0.1,
and component speeds measured by others have been expressed in these terms.
With these assumptions, 1 mas corresponds to a linear distance of 5.8 pc,
and a proper motion of 1 mas yr$^{-1}$ corresponds to an apparent speed of 29$c$.

\section{VLBI OBSERVATIONS}
\label{sec-vlbiobs}

We have observed 3C\,279 at 22 GHz since
the mid 1980's, and at 43 GHz since 1995.  Our first experiments used the Global
VLBI Network which was composed of non-identical antennas at various
observatories.  The data through 1994 were recorded in Mark II mode with 2 MHz
bandwidth, followed by correlation
at the Caltech/JPL Block II Correlator.  The Global Network usually
had three observing sessions per year of which two (at most) included
22 GHz.  During those sessions, seven antennas were able to mount 22
GHz receivers.  In 1991, we added the first antennas in the new NRAO
VLBA.  By the mid 1990's, we used the VLBA alone with
32 MHz bandwidth recorded on tapes correlated at the VLBA Correlator
in Socorro, New Mexico.  We added 43 GHz to our monitoring starting in 1995.  Some of the
later maps were made in ``snapshot'' mode while others were made with
full ($u,v$) tracks.  Most of the full-track observations since 1995 were 
done with alternating scans at 22 and 43 GHz, and sometimes a lower frequency.
Results from this monitoring program prior to 1991 are discussed by U89 and C93.
The data obtained during the more frequent monitoring since
1991 are discussed in this paper; these VLBI observations are listed in Table~\ref{vlbiobs}.
Since the end of our monitoring program in 1997, 3C\,279 has been part of a VLBA
polarization monitoring program described by Marchenko et al. (1999).

\begin{table*}[!t]
\caption{VLBI Observations}
\label{vlbiobs}
\begin{center}
{\scriptsize \begin{tabular}{l l l l c c c c} \tableline \tableline
& \multicolumn{1}{c}{Experiment} & & & \multicolumn{1}{c}{Bandwidth} &
Obs. Time\tablenotemark{c} & Frequencies\tablenotemark{d} & \\ 
\multicolumn{1}{c}{Epoch} & \multicolumn{1}{c}{Name} & VLBA Antennas\tablenotemark{a} 
& Other antennas\tablenotemark{b} & \multicolumn{1}{c}{(MHz)} & 
(minutes) & (GHz) & Polarization \\ \tableline
1991 Jun 24 & GU2B   & Fd,Kp,La,Nl,Pt             & Eb,Gb,Hs,Mc,Nt,On,Ov,Y1 & 2  & 973      & 22     & LCP  \\ 
1992 Jun 14 & GW6B   & Br,Fd,Kp,La,Nl,Ov,Pt       & Eb,Gb,Hs,Mc,Mh,On,Y1    & 2  & 703      & 22     & LCP  \\ 
1992 Nov 10 & GW6C   & Br,Hn,Kp,La,Nl,Ov,Pt       & Gb,Mc,Mh,Nt,On,Y1       & 2  & 485      & 22     & LCP  \\  
1993 Feb 17 & GW008  & Br,Hn,Kp,Ov,Pt,Sc          & Eb,Mc,Mh,Y1             & 2  & 521      & 22     & LCP  \\ 
1993 Nov 8  & BM030  & Br,Hn,Mk,Nl,Ov,Pt          & ...                     & 2  & 115      & 22     & LCP  \\ 
1994 Mar 2  & GW011A & Br,Fd,Hn,Kp,La,Mk,Nl,Ov,Sc & Eb,Gb,Mc,Mh,Nt,On,Y1    & 14 & 755      & 22     & LCP  \\ 
1994 Jun 12 & BM032A & Hn,Mk,Nl,Ov,Pt,Sc          & ...                     & 2  & 96       & 22     & LCP  \\
1994 Sep 21 & GW011B & Br,Fd,Hn,Kp,La,Nl,Ov,Pt,Sc & Y1                      & 14 & 433      & 22     & LCP  \\
1995 Jan 4  & BB025  & All                        & ...                     & 16 & 32       & 22     & Dual \\
1995 Feb 25 & BM038  & All                        & ...                     & 16 & 25, 45   & 22, 43 & Dual \\
1995 Mar 19 & GW013B & All                        & ...                     & 16 & 297, 297 & 22, 43 & Dual \\
1996 Jan 7  & GW013C & All                        & ...                     & 32 & 311, 311 & 22, 43 & LCP  \\
1996 May 4  & BM063  & All                        & ...                     & 32 & 58       &     43 & Dual \\
1996 May 13 & BW026  & All                        & ...                     & 16 & 278      & 22     & Dual \\
1996 Jun 9  & BW026B & All                        & ...                     & 16 & 278, 278 & 22, 43 & Dual \\
1996 Nov 24 & BM072  & All                        & ...                     & 8  & 27       &     43 & Dual \\
1997 Jan 15 & BW026D & All                        & ...                     & 16 & 256, 256 & 22, 43 & Dual \\
1997 Mar 29 & BW031A & Br,Fd,Hn,Kp,La,Mk,Ov,Pt,Sc & ...                     & 16 & 190, 190 & 22, 43 & Dual \\
1997 Jul 16 & BW031B & All                        & ...                     & 16 & 190, 190 & 22, 43 & Dual \\
1997 Nov 16 & BW031C & Br,Fd,Kp,Mk,Nl,Ov,Pt,Sc    & ...                     & 16 & 168, 167 & 22, 43 & Dual \\ \tableline
\end{tabular}}
\end{center}
\tablenotetext{a}{Br = Brewster, WA; Fd = Fort Davis, TX; Hn = Hancock, NH;
Kp = Kitt Peak, AZ; La = Los Alamos, NM; Mk = Mauna Kea, HI; Nl = North Liberty, IA;
Ov = Owens Valley, CA; Pt = Pie Town, NM; Sc = St. Croix, US Virgin Islands.}
\tablenotetext{b}{Antenna locations and sizes are as follows:
Eb = Effelsberg, Germany, 100 m;
Gb = Green Bank, WV, 43 m; Hs = Haystack, MA, 37 m;
Mc = Medicina, Italy, 32 m;
Mh = Mets\"{a}hovi, Finland, 14 m;
Mp = Maryland Point, MD, 26 m;
Nt = Noto, Italy, 32 m;
On = Onsala, Sweden, 20 m;
Ov = Owens Valley, CA, 40m;
Y1 = one antenna of the VLA, Socorro, NM, 25 m.}
\tablenotetext{c}{Two numbers indicate time on source at 22 and 43 GHz respectively.}
\tablenotetext{d}{Lower observed frequencies are not listed here since they
are not discussed in this paper.}
\end{table*}

Images from some of the epochs listed in Table~\ref{vlbiobs}
have appeared in various conference proceedings (e.g. Wehrle et al. 1994; Wehrle et al. 1996; Unwin et al. 1998).
In many cases, we have re-analyzed the original data, and obtained
significantly improved images.  The biggest improvement was in correcting
station-based calibration errors (and deletion of bad data in some cases);
the interactive self-calibration, display, and imaging package DIFMAP (Shepherd, Pearson, \& Taylor 1994)
was the key to realizing these improvements.  For some of the epochs listed in Table~\ref{vlbiobs},
observations at lower frequencies were made as well, but since these observations
contribute little to following source structure changes (because of their lower resolution),
discussion of these observations will be deferred until the discussion of the 
broad-band spectrum in Piner et al. (2000b).  Beginning in 1995 many of our observations recorded
dual circular polarization.  In this paper, we discuss
only the total intensity images formed from these observations.

The data were fringe-fitted in AIPS, then exported to the Caltech
DIFMAP package (Shepherd et al. 1994) for amplitude and phase calibration, editing,
and mapping.  
The 22 GHz data required particular attention to amplitude calibration
because water vapor in the earth's atmosphere absorbs at this
frequency.  Normal self-calibration does not take care of this problem
if it is cloudy at most antennas because there are insufficient
crossing points in the $(u,v)$ plane for sources that are nearly equatorial
(like 3C\,279); moreover, the problem is worse for antennas that are
located far from the mainland array such as Saint Croix (which is
nearly at sea level) or Mauna Kea (which observes at low elevation
angles).  We compared the 22 GHz monitoring data from Mets\"{a}hovi (where
only the data obtained with dry observing
conditions are accepted) with the flux density measured on
the shortest VLBI baselines, and applied an initial scaling factor of
order 1.1 - 1.3 to antenna gains for stations affected by cloudy
weather.  We estimate that 3C\,279 has about 1 Jy in 22 GHz emission
which is too diffuse to be sampled by the shortest spacings in the ($u,v$)
plane.  In most cases, we chose an epoch with good weather to map and
model fit, then used the input model to initiate the
mapping-self-calibration sequence for adjacent epochs with bad weather.
Data from antennas obtained during snow or rain were flagged after we
found that they had significant adverse effects on the images.

The amplitude calibration of the  43 GHz data was compared with the
37 GHz monitoring flux densities from Mets\"{a}hovi.  In nearly all cases, the
fluxes agreed to within 20\%; discrepant antennas were scaled
accordingly.  Self-calibration enables us to make reliable images for the
purpose of tracking changes in the source structure; however, the scale
factors applied render the overall flux scale somewhat uncertain.  This
limits our ability to track the flux density evolution of individual
components at better than about a 10\% level.  
  
\section{VLBI IMAGING RESULTS}
\label{sec-imaging}
Figure~1 shows the eighteen 22 GHz images of 3C\,279 from the epochs listed in Table~\ref{vlbiobs},
and Figure~2 shows the ten 43 GHz images.  The images are shown with uniform weighting 
(uvweight=2,0 in DIFMAP) to maximize the resolution.  Even though this produces an image with
lower dynamic range, the high resolution is important for distinguishing components in the inner milliarcsecond.
The parameters of these images are listed in Table~\ref{imtab}.
Model-fit Gaussian positions are marked with asterisks on the images.  The model-fitting results are discussed
in $\S$~\ref{mfit}.

\begin{sidewaysfigure*}
\plotfiddle{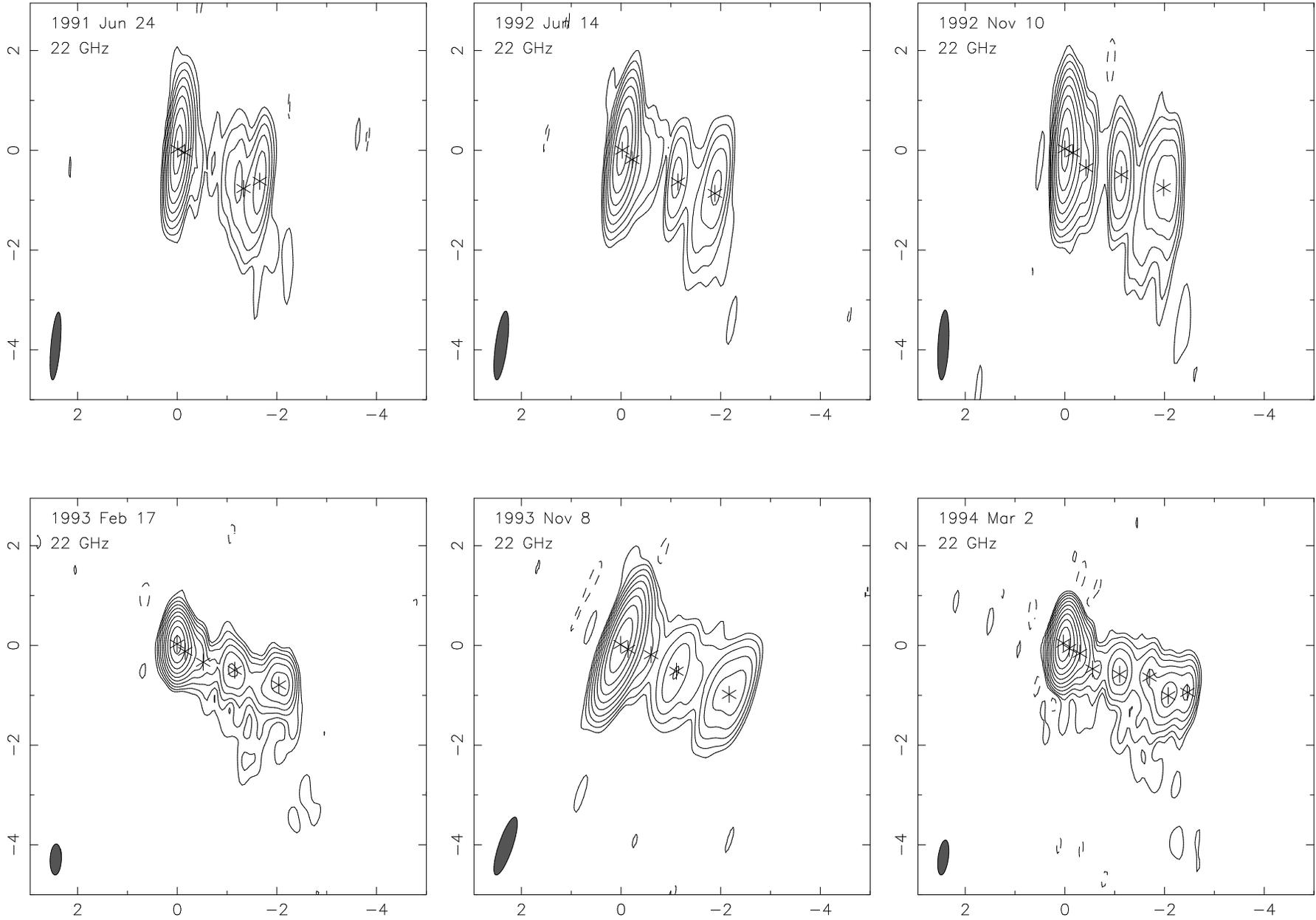}{6.5in}{-90}{90}{90}{-363}{522}
\caption{22 GHz uniformly weighted images of 3C\,279 from the 18 epochs listed in Table~\ref{vlbiobs}.
The axes are labeled in milliarcseconds.  
Parameters of the images are given in Table~\ref{imtab}.
Model-fit Gaussian positions are marked with asterisks.  The model-fit Gaussians are identified
in Table~\ref{mfittab}.}
\end{sidewaysfigure*}

\begin{sidewaysfigure*}
\plotfiddle{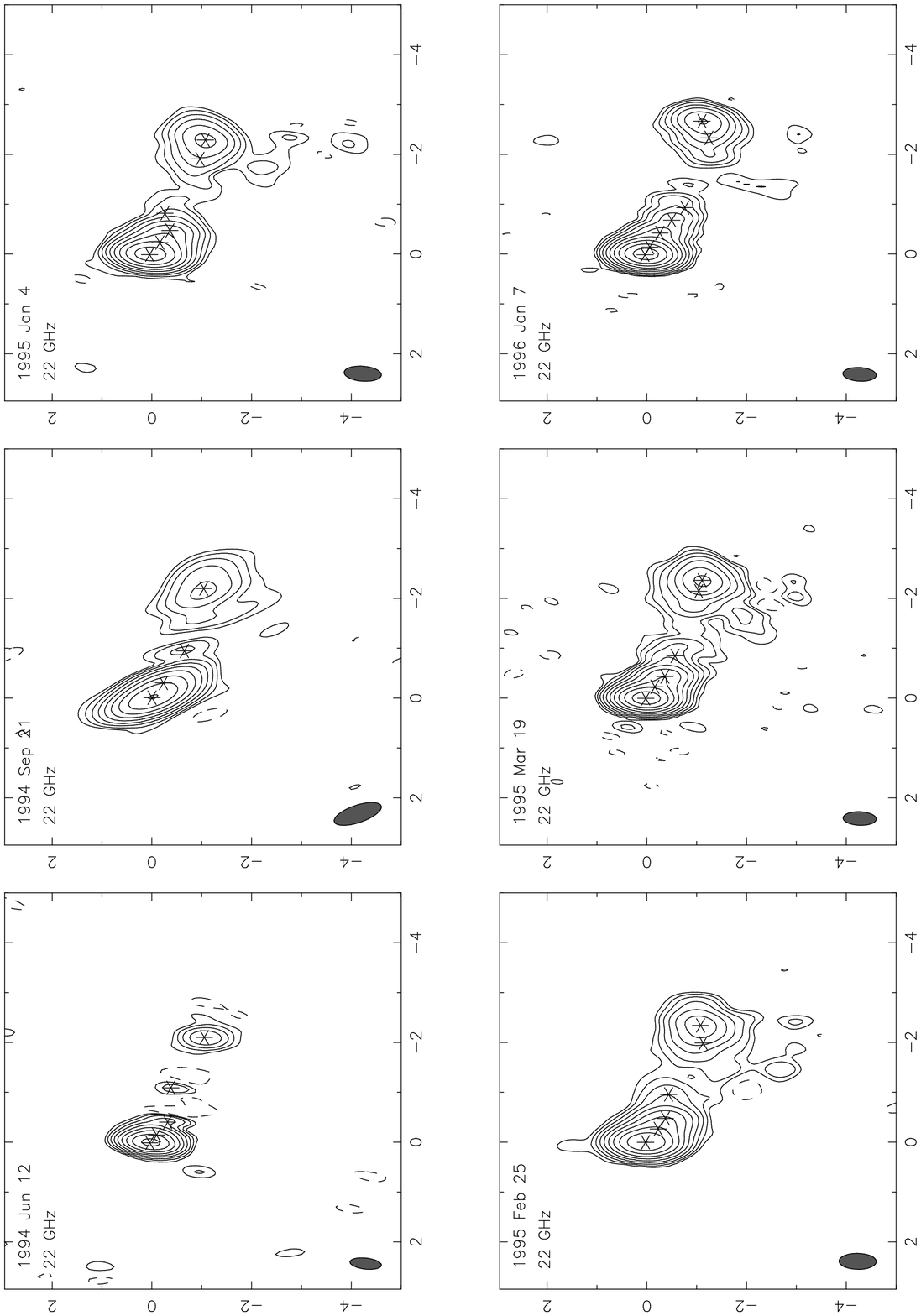}{6.5in}{-90}{90}{90}{-363}{522}
\begin{center}
FIG. 1.---{\em Continued}
\end{center}
\end{sidewaysfigure*}

\begin{sidewaysfigure*}
\plotfiddle{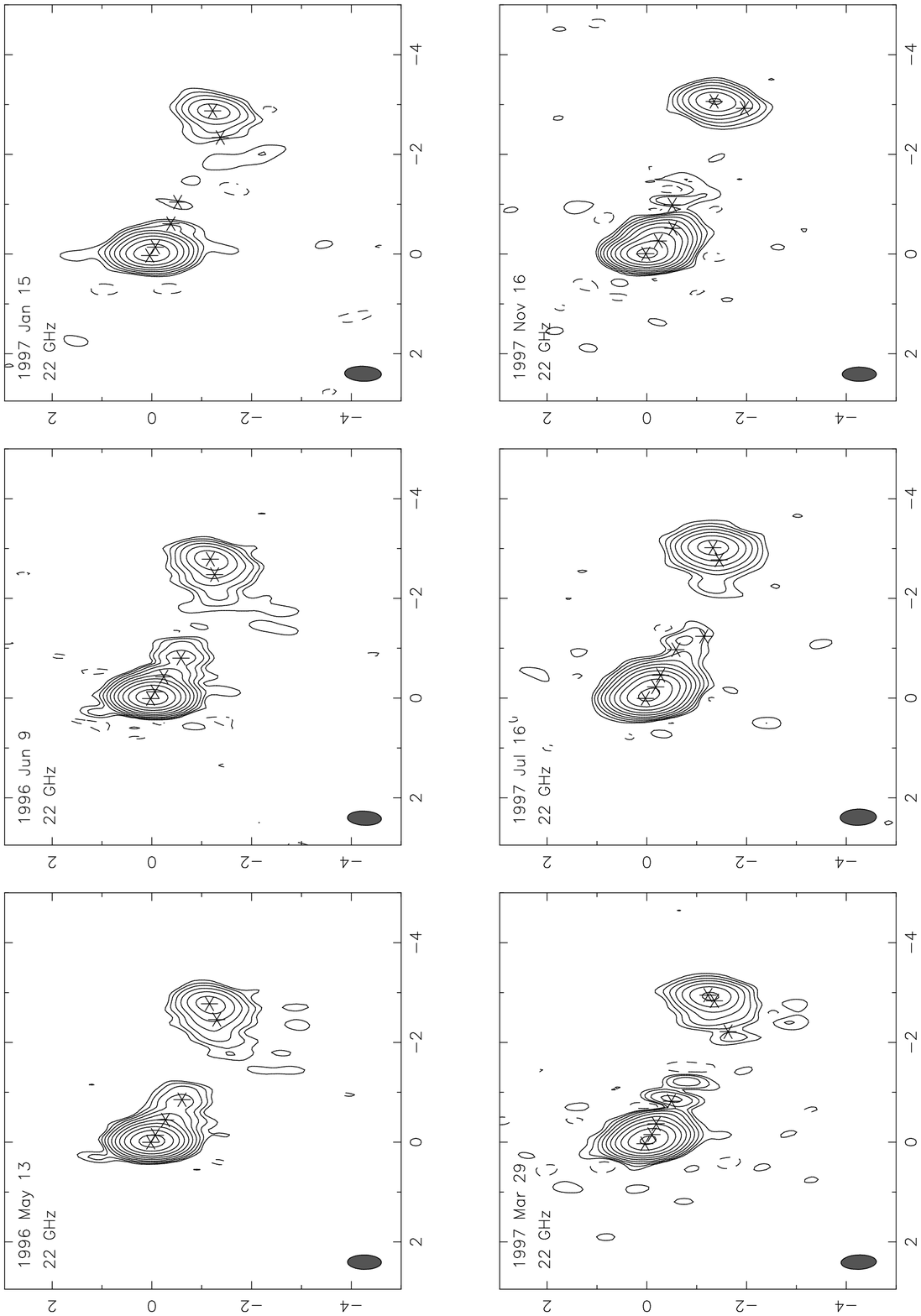}{6.5in}{-90}{90}{90}{-363}{522}
\begin{center}
FIG. 1.---{\em Continued}
\end{center}
\end{sidewaysfigure*}

\begin{sidewaysfigure*}
\plotfiddle{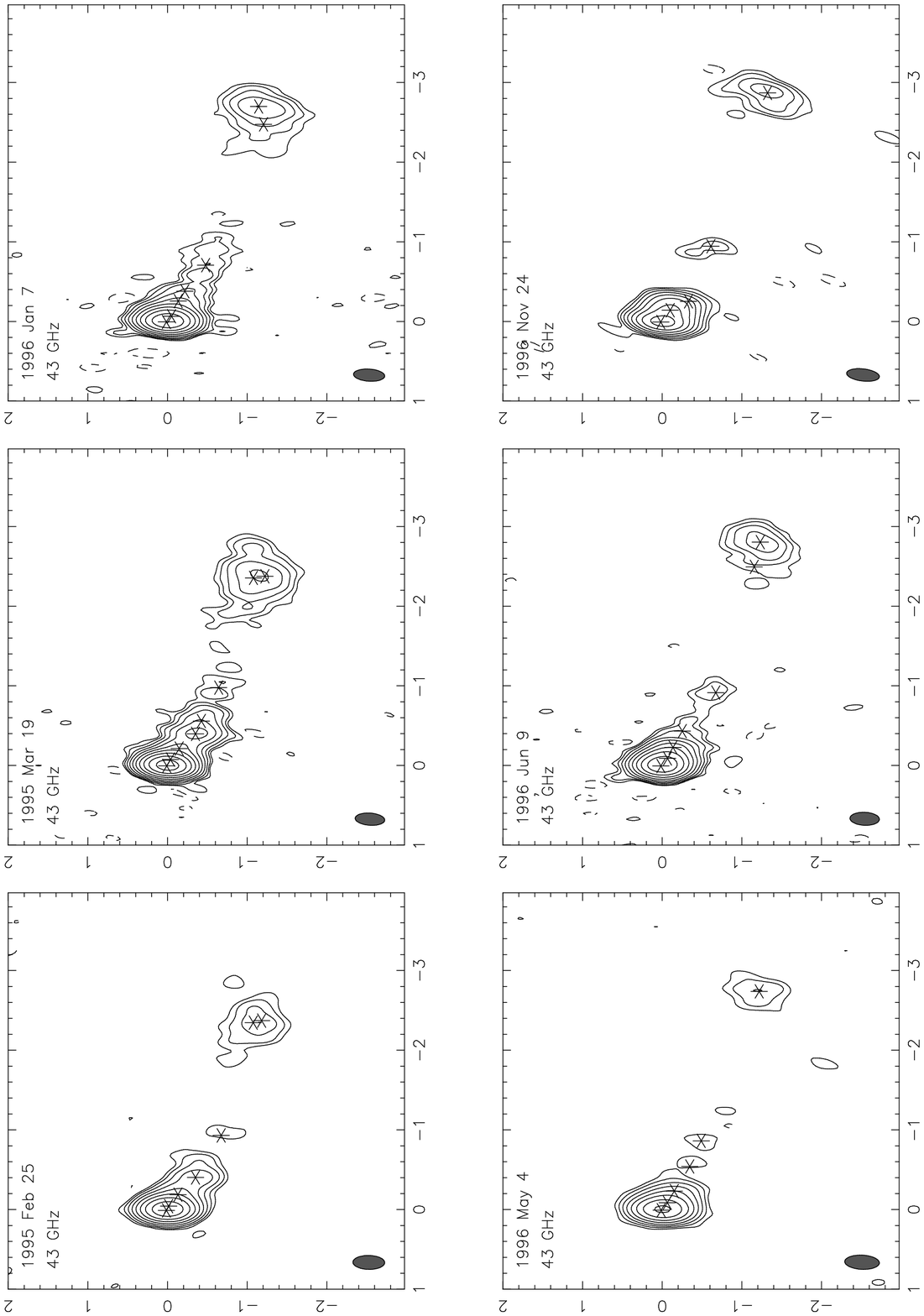}{6.5in}{-90}{90}{90}{-363}{522}
\caption{43 GHz uniformly weighted images of 3C\,279 from the 10 epochs listed in Table~\ref{vlbiobs}.
The axes are labeled in milliarcseconds.  Parameters of the images are given in Table~\ref{imtab}.
Model-fit Gaussian positions are marked with asterisks.  The model-fit Gaussians are identified
in Table~\ref{mfittab}.}
\end{sidewaysfigure*}

\begin{sidewaysfigure*}
\plotfiddle{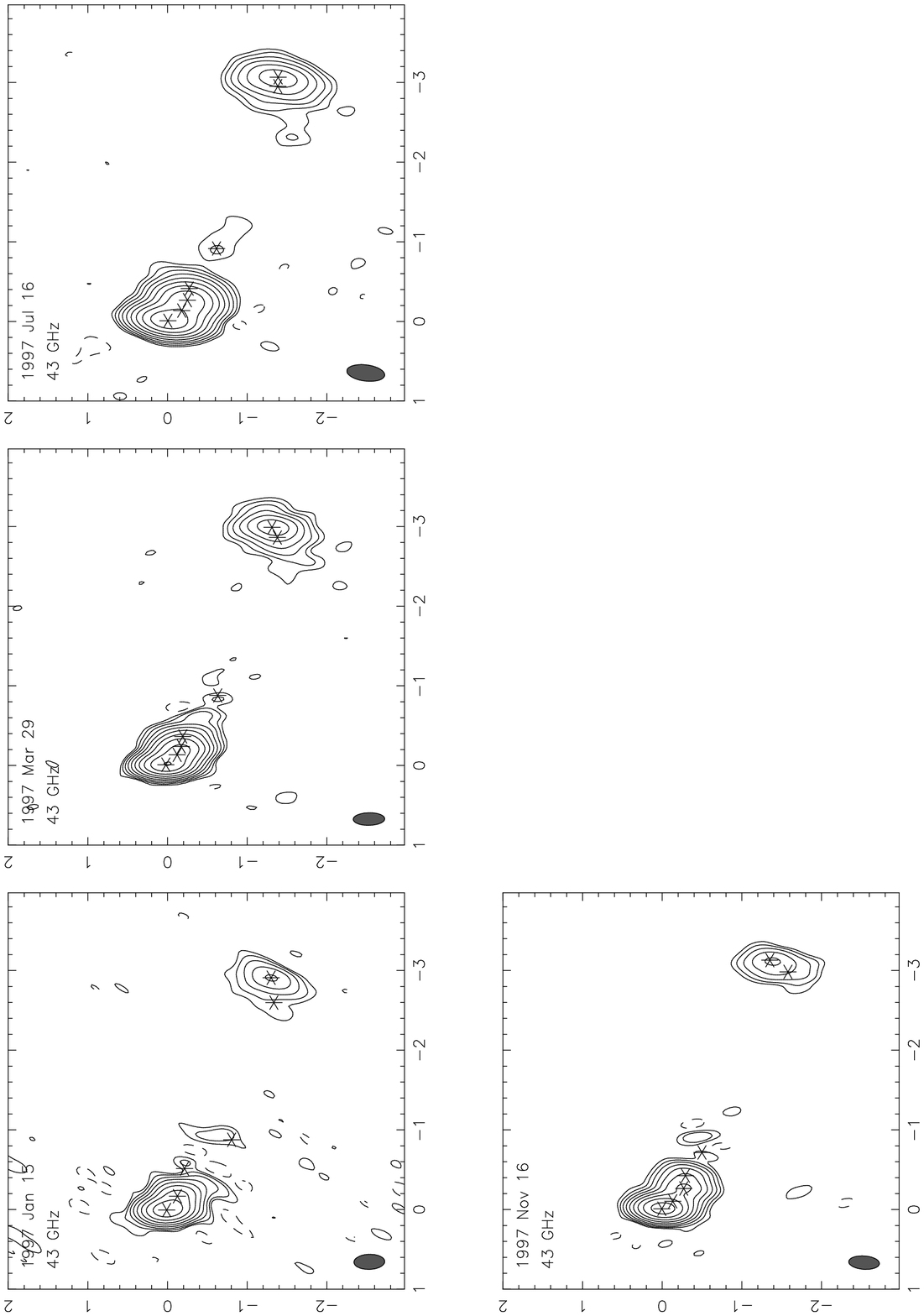}{6.5in}{-90}{90}{90}{-363}{522}
\begin{center}
FIG. 2.---{\em Continued}
\end{center}
\end{sidewaysfigure*}

\begin{table*}[!t]
\caption{Parameters of the Images}
\label{imtab}
\begin{center}
{\small \begin{tabular}{l c l c c c c c} \tableline \tableline
& & & Total & CLEAN & Peak & Lowest & Contours\tablenotemark{d} \\ 
& Frequency & & Flux\tablenotemark{b} & Flux & Flux & Contour\tablenotemark{c} & (multiples of \\ 
\multicolumn{1}{c}{Epoch} & (GHz) & \multicolumn{1}{c}{Beam\tablenotemark{a}} & (Jy) & (Jy) &
(Jy beam$^{-1}$) & (mJy beam$^{-1}$) & lowest contour) \\ \tableline
1991 Jun 24 & 22 & 1.36,0.19,$-$5.0  & 15.7 & 15.3 & 9.4  & 50.5  & 1...2$^{7}$  \\ 
1992 Jun 14 & 22 & 1.39,0.24,$-$7.6  & 14.1 & 14.0 & 9.3  & 51.4  & 1...2$^{7}$  \\ 
1992 Nov 10 & 22 & 1.40,0.21,$-$3.1  & 15.2 & 15.3 & 9.6  & 28.6  & 1...2$^{8}$  \\ 
1993 Feb 17 & 22 & 0.62,0.23,$-$2.3  & 15.8 & 16.2 & 10.6 & 16.2  & 1...2$^{9}$  \\ 
1993 Nov 8  & 22 & 1.22,0.28,$-$18.7 & 19.9 & 20.2 & 15.3 & 31.9  & 1...2$^{8}$  \\ 
1994 Mar 2  & 22 & 0.71,0.21,$-$7.7  & 20.5 & 20.0 & 12.9 & 13.1  & 1...2$^{9}$  \\  
1994 Jun 12 & 22 & 0.63,0.22,$-$6.4  & ...  & 19.7 & 15.1 & 47.0  & 1...2$^{8}$  \\ 
1994 Sep 21 & 22 & 0.99,0.35,18.7    & ...  & 20.1 & 14.9 & 56.7  & 1...2$^{8}$  \\ 
1995 Jan 4  & 22 & 0.75,0.30,$-$4.3  & 18.3 & 18.6 & 12.1 & 29.4  & 1...2$^{8}$  \\ 
1995 Feb 25 & 22 & 0.75,0.32,$-$0.9  & 18.6 & 18.5 & 12.3 & 23.9  & 1...2$^{9}$  \\ 
1995 Mar 19 & 22 & 0.67,0.27,$-$2.1  & 19.2 & 18.9 & 12.3 & 14.7  & 1...2$^{9}$  \\ 
1996 Jan 7  & 22 & 0.67,0.28,$-$3.3  & 20.4 & 19.7 & 14.1 & 18.6  & 1...2$^{9}$  \\ 
1996 May 13 & 22 & 0.68,0.29,$-$0.6  & 24.1 & 23.5 & 17.6 & 23.3  & 1...2$^{9}$  \\ 
1996 Jun 9  & 22 & 0.68,0.29,$-$2.6  & 28.4 & 27.7 & 20.9 & 22.4  & 1...2$^{9}$  \\ 
1997 Jan 15 & 22 & 0.74,0.30,$-$2.0  & 24.5 & 24.9 & 17.6 & 73.0  & 1...2$^{7}$  \\ 
1997 Mar 29 & 22 & 0.71,0.29,2.1     & 24.8 & 23.8 & 14.9 & 25.7  & 1...2$^{9}$  \\ 
1997 Jul 16 & 22 & 0.72,0.33,1.4     & 27.6 & 25.1 & 14.4 & 22.9  & 1...2$^{9}$  \\ 
1997 Nov 16 & 22 & 0.68,0.29,0.1     & 35.4 & 34.7 & 21.9 & 33.6  & 1...2$^{9}$  \\ 
1995 Feb 25 & 43 & 0.40,0.18,$-$0.9  & 19.5 & 18.4 & 13.5 & 55.6  & 1...2$^{7}$  \\ 
1995 Mar 19 & 43 & 0.37,0.15,$-$4.0  & 19.6 & 18.6 & 13.6 & 18.0  & 1...2$^{9}$  \\ 
1996 Jan 7  & 43 & 0.39,0.15,$-$3.7  & 20.8 & 19.9 & 14.6 & 28.9  & 1...2$^{8}$  \\ 
1996 May 4  & 43 & 0.44,0.18,$-$2.7  & 22.4 & 22.5 & 17.1 & 117   & 1...2$^{7}$  \\ 
1996 Jun 9  & 43 & 0.37,0.16,$-$2.5  & 26.4 & 24.6 & 17.4 & 35.1  & 1...2$^{8}$  \\ 
1996 Nov 24 & 43 & 0.41,0.15,$-$6.7  & 21.3 & 22.3 & 12.7 & 69.2  & 1...2$^{7}$  \\ 
1997 Jan 15 & 43 & 0.39,0.19,1.8     & 21.9 & 21.5 & 14.2 & 60.7  & 1...2$^{7}$  \\ 
1997 Mar 29 & 43 & 0.40,0.16,1.1     & 21.8 & 19.6 & 9.0  & 16.9  & 1...2$^{9}$  \\ 
1997 Jul 16 & 43 & 0.48,0.20,$-$7.2  & 20.9 & 20.0 & 10.4 & 21.6  & 1...2$^{8}$  \\ 
1997 Nov 16 & 43 & 0.39,0.17,$-$3.9  & 33.7 & 32.7 & 21.0 & 55.5  & 1...2$^{8}$  \\ \tableline
\end{tabular}}
\end{center}
\tablenotetext{a}{Numbers given for the beam are the FWHMs of the major
and minor axes in mas, and the position angle of the major axis in degrees.
The beam has been synthesized using uniform weighting.}
\tablenotetext{b}{Single-dish flux from Mets\"{a}hovi at 22 or 37 GHz.  Absent entries indicate
periods of no observing.}
\tablenotetext{c}{The lowest contour is set to be three times the rms noise
in the full image.}
\tablenotetext{d}{Contour levels are represented by the geometric series 1...2$^{n}$,
e.g. for $n=5$ the contour levels would be $\pm$1,2,4,8,16,32.}
\end{table*}

The parsec-scale morphology of 3C\,279 during the years 1991-1997 
consists of the bright compact core, a
bright secondary component at a position angle of $-114\arcdeg$ which moves outward from about 2 mas to
about 3.5 mas from the core during the observed time range, and 
an inner jet that extends from the core out to about 1 mas.
We identify the bright secondary component with
the component C4 seen previously in our monitoring (U89; C93), 
and subsequently by many other authors.
In the earlier images, C4 is connected to the inner jet emission, but in the later images
it is clearly separated, with a gap in emission between the 1 mas point and C4.
The region interior to 1 mas is complex, with multiple components forming, moving out,
and fading on timescales of several years.  Component C4 is resolved and has significant 
internal structure, with multiple Gaussians often required to model it (see $\S$~\ref{mfit}).
The higher dynamic range images (e.g. the 22 GHz images from 1993 Feb 17 and 1994 Mar 2)
show extended, diffuse emission to the southwest of the main jet.  This extended, diffuse
emission has a position angle of about $-140\arcdeg$ to $-150\arcdeg$, similar to that of the larger-scale
VLBI jet seen at lower frequencies (Piner et al. 2000a), and the kiloparsec-scale jet
seen with the VLA and MERLIN (de Pater \& Perley 1983; Pilbratt, Booth, \& Porcas 1987; Akujor et al. 1994).  
No counterjet is detected, and the limit placed on the
jet/counterjet brightness ratio is about 100:1 at the distance of C4.

\section{MOTION OF JET COMPONENTS}
\label{sec-slmotion}

\subsection{Identification of Components by Model Fitting}
\label{mfit}
We used the modelfit routine in DIFMAP to fit Gaussians to the visibility
data for each epoch.  These Gaussian models are listed in Table~\ref{mfittab}.
Our procedure was to replace all CLEAN components with a collection of circular Gaussians,
letting the circular Gaussians become elliptical if required to fit the visibility data
and residual map.  All regions of the jet could be successfully modeled with circular
Gaussians with the exception of the resolved jet component C4, which required one or more elliptical
Gaussians at some epochs.  Use of circular Gaussians has the advantage that
it prevents representation of one or more jet components by a single, long, elliptical Gaussian.

\begin{table*}
\caption{Gaussian Models}
\label{mfittab}
\begin{center}
{\small \begin{tabular}{l c c r r r r r r} \tableline \tableline
& Frequency & Component\tablenotemark{a}
& \multicolumn{1}{c}{$S$\tablenotemark{b}} & \multicolumn{1}{c}{$r$\tablenotemark{c}} &
\multicolumn{1}{c}{PA\tablenotemark{c}} &
\multicolumn{1}{c}{$a$\tablenotemark{d}} &
& \multicolumn{1}{c}{$\Phi$\tablenotemark{e}} \\ 
\multicolumn{1}{c}{Epoch} & (GHz) & ID
& \multicolumn{1}{c}{(Jy)} & \multicolumn{1}{c}{(mas)} &
\multicolumn{1}{c}{(deg)} 
& \multicolumn{1}{c}{(mas)} & \multicolumn{1}{c}{$b/a$} & \multicolumn{1}{c}{(deg)} \\ \tableline
1991 Jun 24 & 22 & Core  &  9.89 &  ... &    ... & 0.10 & 1.00 &    ... \\ 
            &    & C5a   &  1.82 & 0.15 & -114.9 & 0.25 & 1.00 &    ... \\ 
            &    & C5    &  2.18 & 1.54 & -120.8 & 0.87 & 1.00 &    ... \\ 
            &    & C4    &  1.42 & 1.77 & -111.3 & 0.14 & 1.00 &    ... \\
1992 Jun 14 & 22 & Core  &  9.82 &  ... &    ... & 0.11 & 1.00 &    ... \\ 
            &    & C5a   &  1.29 & 0.28 & -133.6 & 0.17 & 1.00 &    ... \\ 
            &    & C5    &  1.33 & 1.29 & -119.6 & 0.95 & 1.00 &    ... \\ 
            &    & C4    &  1.52 & 2.05 & -115.0 & 0.27 & 1.00 &    ... \\
1992 Nov 10 & 22 & Core  &  8.93 &  ... &    ... & 0.03 & 1.00 &    ... \\ 
            &    & C6    &  2.52 & 0.18 & -117.0 & 0.14 & 1.00 &    ... \\ 
            &    & C5a   &  0.60 & 0.57 & -131.9 & 0.22 & 1.00 &    ... \\ 
            &    & C5    &  0.98 & 1.25 & -114.7 & 0.27 & 1.00 &    ... \\ 
            &    & C4    &  2.14 & 2.13 & -111.5 & 0.46 & 1.00 &    ... \\
1993 Feb 17 & 22 & Core  & 10.86 &  ... &    ... & 0.09 & 1.00 &    ... \\ 
            &    & C6    &  1.73 & 0.22 & -132.1 & 0.19 & 1.00 &    ... \\ 
            &    & C5a   &  0.57 & 0.64 & -124.8 & 0.35 & 1.00 &    ... \\ 
            &    & C5    &  1.11 & 1.27 & -114.4 & 0.33 & 1.00 &    ... \\ 
            &    & C4    &  1.80 & 2.20 & -111.9 & 0.44 & 1.00 &    ... \\
1993 Nov 8  & 22 & Core  & 14.14 &  ... &    ... & 0.07 & 1.00 &    ... \\ 
            &    & C6/7  &  3.57 & 0.18 & -123.2 & 0.19 & 1.00 &    ... \\ 
            &    & C5a   &  0.21 & 0.65 & -108.6 & 0.00 & 1.00 &    ... \\ 
            &    & C5    &  0.95 & 1.25 & -115.6 & 0.39 & 1.00 &    ... \\ 
            &    & C4    &  1.87 & 2.40 & -114.7 & 0.53 & 1.00 &    ... \\
1994 Mar 2  & 22 & Core  & 12.31 &  ... &    ... & 0.08 & 1.00 &    ... \\ 
            &    & C7    &  3.80 & 0.16 & -132.0 & 0.11 & 1.00 &    ... \\ 
            &    & C6    &  0.82 & 0.39 & -121.6 & 0.11 & 1.00 &    ... \\ 
            &    & C5a   &  0.23 & 0.78 & -131.5 & 0.24 & 1.00 &    ... \\ 
            &    & C5    &  0.62 & 1.29 & -119.0 & 0.34 & 1.00 &    ... \\ 
            &    & C4    &  0.38 & 1.86 & -111.5 & 0.27 & 1.00 &    ... \\ 
            &    & C4*   &  1.26 & 2.35 & -116.6 & 0.33 & 1.00 &    ... \\ 
            &    & C4    &  0.45 & 2.68 & -111.6 & 0.07 & 1.00 &    ... \\ 
1994 Jun 12 & 22 & Core  & 15.53 &  ... &    ... & 0.10 & 1.00 &    ... \\ 
            &    & C7    &  3.36 & 0.20 & -131.2 & 0.15 & 1.00 &    ... \\ 
            &    & C6    &  0.32 & 0.55 & -130.9 & 0.00 & 1.00 &    ... \\ 
            &    & C5    &  0.22 & 1.18 & -110.9 & 0.17 & 1.00 &    ... \\ 
            &    & C4    &  1.20 & 2.38 & -117.3 & 0.28 & 1.00 &    ... \\
1994 Sep 21 & 22 & Core  & 14.54 &  ... &    ... & 0.10 & 1.00 &    ... \\ 
            &    & C6/7  &  2.92 & 0.37 & -127.5 & 0.13 & 1.00 &    ... \\ 
            &    & C5    &  0.51 & 1.14 & -124.6 & 0.90 & 1.00 &    ... \\ 
            &    & C4    &  2.24 & 2.42 & -115.4 & 0.59 & 1.00 &    ... \\
1995 Jan 4  & 22 & Core  & 12.07 &  ... &    ... & 0.07 & 1.00 &    ... \\ 
            &    & C7    &  2.66 & 0.32 & -130.6 & 0.12 & 1.00 &    ... \\ 
            &    & C6    &  1.11 & 0.63 & -130.1 & 0.00 & 1.00 &    ... \\ 
            &    & C5a   &  0.23 & 0.88 & -110.3 & 0.37 & 1.00 &    ... \\ 
            &    & C4    &  0.84 & 2.17 & -117.8 & 1.04 & 1.00 &    ... \\ 
            &    & C4*   &  1.73 & 2.56 & -115.9 & 0.37 & 1.00 &    ... \\
1995 Feb 25 & 22 & Core  & 12.55 &  ... &    ... & 0.09 & 1.00 &    ... \\ 
            &    & C7    &  1.76 & 0.37 & -133.5 & 0.00 & 1.00 &    ... \\ 
            &    & C6    &  1.39 & 0.62 & -129.9 & 0.18 & 1.00 &    ... \\ 
            &    & C5a   &  0.27 & 1.07 & -115.7 & 0.39 & 1.00 &    ... \\ 
            &    & C4    &  0.78 & 2.30 & -120.1 & 0.94 & 1.00 &    ... \\ 
            &    & C4*   &  1.75 & 2.59 & -115.2 & 0.38 & 1.00 &    ... \\
1995 Mar 19 & 22 & Core  & 12.67 &  ... &    ... & 0.08 & 1.00 &    ... \\ 
            &    & C7    &  1.47 & 0.29 & -128.1 & 0.07 & 1.00 &    ... \\ 
            &    & C6    &  1.81 & 0.58 & -131.1 & 0.17 & 1.00 &    ... \\ 
            &    & C5a   &  0.43 & 1.03 & -124.5 & 0.50 & 1.00 &    ... \\ 
            &    & C4    &  0.97 & 2.40 & -116.3 & 0.86 & 1.00 &    ... \\ 
            &    & C4*   &  1.58 & 2.62 & -115.5 & 0.34 & 1.00 &    ... \\
\end{tabular}}
\end{center}
\end{table*}

\begin{table*}
\begin{center}
TABLE 3---{\em Continued} \\
\vspace{0.125in}
{\small \begin{tabular}{l c c r r r r r r} \tableline \tableline
& Frequency & Component\tablenotemark{a}
& \multicolumn{1}{c}{$S$\tablenotemark{b}} & \multicolumn{1}{c}{$r$\tablenotemark{c}} &
\multicolumn{1}{c}{PA\tablenotemark{c}} &
\multicolumn{1}{c}{$a$\tablenotemark{d}} &
& \multicolumn{1}{c}{$\Phi$\tablenotemark{e}} \\ 
\multicolumn{1}{c}{Epoch} & (GHz) & ID
& \multicolumn{1}{c}{(Jy)} & \multicolumn{1}{c}{(mas)} &
\multicolumn{1}{c}{(deg)} 
& \multicolumn{1}{c}{(mas)} & \multicolumn{1}{c}{$b/a$} & \multicolumn{1}{c}{(deg)} \\ \tableline
1996 Jan 7  & 22 & Core  & 13.19 &  ... &    ... & 0.06 & 1.00 &    ... \\ 
            &    & C7a/8 &  2.86 & 0.18 & -122.4 & 0.14 & 1.00 &    ... \\ 
            &    & C7    &  0.60 & 0.53 & -124.4 & 0.00 & 1.00 &    ... \\ 
            &    & C6    &  0.49 & 0.88 & -127.7 & 0.19 & 1.00 &    ... \\ 
            &    & C5    &  0.16 & 1.24 & -130.2 & 0.08 & 1.00 &    ... \\ 
            &    & C4    &  0.92 & 2.67 & -118.7 & 0.64 & 1.00 &    ... \\ 
            &    & C4*   &  1.47 & 2.91 & -113.1 & 0.22 & 1.00 &    ... \\ 
1996 May 13 & 22 & Core  & 15.37 &  ... &    ... & 0.06 & 1.00 &    ... \\ 
            &    & C7a/8 &  4.92 & 0.17 & -123.1 & 0.11 & 1.00 &    ... \\ 
            &    & C7    &  0.59 & 0.54 & -123.6 & 0.14 & 1.00 &    ... \\ 
            &    & C6    &  0.39 & 1.06 & -126.3 & 0.32 & 1.00 &    ... \\ 
            &    & C4    &  0.85 & 2.80 & -118.3 & 0.64 & 1.00 &    ... \\ 
            &    & C4*   &  1.35 & 3.02 & -113.0 & 0.22 & 1.00 &    ... \\
1996 Jun 9  & 22 & Core  & 17.25 &  ... &    ... & 0.04 & 1.00 &    ... \\ 
            &    & C7a/8 &  7.00 & 0.17 & -123.0 & 0.10 & 1.00 &    ... \\ 
            &    & C7    &  0.54 & 0.51 & -121.3 & 0.00 & 1.00 &    ... \\ 
            &    & C6    &  0.50 & 1.02 & -127.0 & 0.38 & 1.00 &    ... \\ 
            &    & C4    &  0.92 & 2.80 & -117.3 & 0.67 & 1.00 &    ... \\ 
            &    & C4*   &  1.51 & 3.04 & -113.2 & 0.36 & 0.26 &  -41.4 \\
1997 Jan 15 & 22 & Core  & 13.44 &  ... &    ... & 0.00 & 1.00 &    ... \\ 
            &    & C8    &  8.23 & 0.20 & -124.7 & 0.12 & 1.00 &    ... \\ 
            &    & C7    &  0.27 & 0.77 & -123.7 & 0.25 & 1.00 &    ... \\ 
            &    & C5    &  0.10 & 1.21 & -117.7 & 0.08 & 1.00 &    ... \\ 
            &    & C4    &  0.73 & 2.76 & -120.9 & 0.92 & 1.00 &    ... \\ 
            &    & C4*   &  2.29 & 3.17 & -113.5 & 0.29 & 0.34 &  -47.9 \\
1997 Mar 29 & 22 & Core  & 10.62 &  ... &    ... & 0.00 & 1.00 &    ... \\ 
            &    & C8/9  &  9.18 & 0.23 & -127.7 & 0.10 & 1.00 &    ... \\ 
            &    & C7a   &  0.80 & 0.46 & -119.6 & 0.00 & 1.00 &    ... \\ 
            &    & C7    &  0.32 & 1.00 & -121.9 & 0.50 & 1.00 &    ... \\ 
            &    & C4    &  0.26 & 2.80 & -126.6 & 0.74 & 1.00 &    ... \\ 
            &    & C4    &  1.09 & 3.19 & -115.9 & 0.36 & 1.00 &    ... \\ 
            &    & C4*   &  1.60 & 3.24 & -113.0 & 0.30 & 0.14 &  -26.1 \\ 
1997 Jul 16 & 22 & Core  & 11.37 &  ... &    ... & 0.04 & 1.00 &    ... \\ 
            &    & C8/9  &  9.62 & 0.31 & -131.9 & 0.17 & 1.00 &    ... \\ 
            &    & C7a   &  0.73 & 0.56 & -123.3 & 0.00 & 1.00 &    ... \\ 
            &    & C5    &  0.10 & 1.15 & -122.1 & 0.15 & 1.00 &    ... \\ 
            &    & C5a?  &  0.07 & 1.71 & -133.2 & 0.11 & 1.00 &    ... \\ 
            &    & C4    &  0.48 & 3.15 & -118.1 & 0.78 & 1.00 &    ... \\ 
            &    & C4*   &  2.72 & 3.31 & -114.1 & 0.32 & 0.57 &  -23.4 \\ 
1997 Nov 16 & 22 & Core  & 21.36 &  ... &    ... & 0.08 & 1.00 &    ... \\ 
            &    & C8/9  &  9.46 & 0.36 & -134.3 & 0.26 & 1.00 &    ... \\ 
            &    & C7/7a &  0.58 & 0.75 & -136.6 & 0.00 & 1.00 &    ... \\ 
            &    & C5    &  0.18 & 1.12 & -118.3 & 0.00 & 1.00 &    ... \\ 
            &    & C4*   &  3.10 & 3.35 & -114.1 & 0.25 & 0.78 &  -56.9 \\ 
            &    & C4    &  0.33 & 3.53 & -124.1 & 0.00 & 1.00 &    ... \\ 
1995 Feb 25 & 43 & Core  &  9.47 &  ... &    ... & 0.00 & 1.00 &    ... \\ 
            &    & C7a   &  5.34 & 0.06 & -117.5 & 0.08 & 1.00 &    ... \\ 
            &    & C7    &  0.92 & 0.25 & -126.3 & 0.10 & 1.00 &    ... \\ 
            &    & C6    &  1.27 & 0.55 & -131.4 & 0.21 & 1.00 &    ... \\ 
            &    & C5    &  0.20 & 1.17 & -126.2 & 0.20 & 1.00 &    ... \\ 
            &    & C4*   &  1.03 & 2.60 & -114.8 & 0.57 & 1.00 &    ... \\ 
            &    & C4    &  0.28 & 2.67 & -116.7 & 0.11 & 1.00 &    ... \\
1995 Mar 19 & 43 & Core  & 12.57 &  ... &    ... & 0.03 & 1.00 &    ... \\ 
            &    & C7a   &  2.78 & 0.09 & -121.3 & 0.05 & 1.00 &    ... \\ 
            &    & C7    &  0.67 & 0.27 & -125.8 & 0.11 & 1.00 &    ... \\ 
            &    & C6    &  0.85 & 0.54 & -131.9 & 0.10 & 1.00 &    ... \\ 
            &    & C5a   &  0.42 & 0.71 & -127.7 & 0.19 & 1.00 &    ... \\ 
            &    & C5    &  0.17 & 1.18 & -123.6 & 0.37 & 1.00 &    ... \\ 
            &    & C4*   &  1.08 & 2.60 & -114.8 & 0.51 & 1.00 &    ... \\ 
            &    & C4    &  0.24 & 2.68 & -117.4 & 0.14 & 1.00 &    ... \\
\end{tabular}}
\end{center}
\end{table*}

\begin{table*}
\begin{center}
TABLE 3---{\em Continued} \\
\vspace{0.125in}
{\small \begin{tabular}{l c c r r r r r r} \tableline \tableline
& Frequency & Component\tablenotemark{a}
& \multicolumn{1}{c}{$S$\tablenotemark{b}} & \multicolumn{1}{c}{$r$\tablenotemark{c}} &
\multicolumn{1}{c}{PA\tablenotemark{c}} &
\multicolumn{1}{c}{$a$\tablenotemark{d}} &
& \multicolumn{1}{c}{$\Phi$\tablenotemark{e}} \\ 
\multicolumn{1}{c}{Epoch} & (GHz) & ID
& \multicolumn{1}{c}{(Jy)} & \multicolumn{1}{c}{(mas)} &
\multicolumn{1}{c}{(deg)} 
& \multicolumn{1}{c}{(mas)} & \multicolumn{1}{c}{$b/a$} & \multicolumn{1}{c}{(deg)} \\ \tableline
1996 Jan 7  & 43 & Core  & 12.43 &  ... &    ... & 0.04 & 1.00 &    ... \\ 
            &    & C8    &  4.89 & 0.10 & -132.5 & 0.08 & 1.00 &    ... \\ 
            &    & C7a   &  0.69 & 0.30 & -119.7 & 0.15 & 1.00 &    ... \\ 
            &    & C7    &  0.11 & 0.45 & -121.1 & 0.00 & 1.00 &    ... \\ 
            &    & C6    &  0.49 & 0.87 & -124.7 & 0.44 & 1.00 &    ... \\ 
            &    & C4    &  0.59 & 2.76 & -116.2 & 0.62 & 1.00 &    ... \\ 
            &    & C4*   &  0.88 & 2.94 & -113.2 & 0.45 & 0.40 &  -15.0 \\
1996 May 4  & 43 & Core  & 15.01 &  ... &    ... & 0.03 & 1.00 &    ... \\ 
            &    & C8    &  4.63 & 0.12 & -128.9 & 0.05 & 1.00 &    ... \\ 
            &    & C7a   &  1.32 & 0.29 & -124.4 & 0.06 & 1.00 &    ... \\ 
            &    & C7    &  0.19 & 0.66 & -123.2 & 0.09 & 1.00 &    ... \\ 
            &    & C6    &  0.23 & 1.01 & -120.0 & 0.06 & 1.00 &    ... \\ 
            &    & C4    &  1.14 & 3.01 & -114.1 & 0.31 & 1.00 &    ... \\
1996 Jun 9  & 43 & Core  & 15.67 &  ... &    ... & 0.05 & 1.00 &    ... \\ 
            &    & C8    &  6.16 & 0.13 & -131.1 & 0.08 & 1.00 &    ... \\ 
            &    & C7a   &  1.21 & 0.28 & -123.2 & 0.05 & 1.00 &    ... \\ 
            &    & C7    &  0.25 & 0.51 & -121.6 & 0.22 & 1.00 &    ... \\ 
            &    & C5    &  0.19 & 1.15 & -126.8 & 0.14 & 1.00 &    ... \\ 
            &    & C4    &  0.29 & 2.76 & -115.1 & 0.86 & 0.46 &  -34.0 \\ 
            &    & C4*   &  0.98 & 3.08 & -114.0 & 0.40 & 0.42 &  -35.8 \\
1996 Nov 24 & 43 & Core  & 11.44 &  ... &    ... & 0.05 & 1.00 &    ... \\ 
            &    & C8    &  8.84 & 0.19 & -128.6 & 0.20 & 1.00 &    ... \\ 
            &    & C7a   &  0.37 & 0.43 & -143.2 & 0.00 & 1.00 &    ... \\ 
            &    & C5    &  0.24 & 1.14 & -123.5 & 0.20 & 1.00 &    ... \\ 
            &    & C4    &  1.78 & 3.17 & -115.0 & 0.22 & 1.00 &    ... \\ 
1997 Jan 15 & 43 & Core  & 13.34 &  ... &    ... & 0.03 & 1.00 &    ... \\ 
            &    & C8    &  6.04 & 0.22 & -127.7 & 0.12 & 1.00 &    ... \\ 
            &    & C7/7a &  0.16 & 0.57 & -113.3 & 0.00 & 1.00 &    ... \\ 
            &    & C5    &  0.16 & 1.20 & -132.8 & 0.04 & 1.00 &    ... \\ 
            &    & C4    &  0.20 & 2.93 & -117.3 & 0.00 & 1.00 &    ... \\ 
            &    & C4*   &  1.68 & 3.20 & -114.2 & 0.38 & 0.10 &  -34.4 \\
1997 Mar 29 & 43 & Core  &  8.80 &  ... &    ... & 0.06 & 1.00 &    ... \\ 
            &    & C9    &  4.05 & 0.19 & -138.7 & 0.07 & 1.00 &    ... \\ 
            &    & C8    &  4.02 & 0.30 & -130.1 & 0.08 & 1.00 &    ... \\ 
            &    & C7a   &  0.81 & 0.41 & -121.0 & 0.15 & 1.00 &    ... \\ 
            &    & C5    &  0.18 & 1.09 & -126.9 & 0.30 & 1.00 &    ... \\ 
            &    & C4    &  0.64 & 3.18 & -116.2 & 0.67 & 1.00 &    ... \\ 
            &    & C4*   &  1.28 & 3.26 & -114.1 & 0.25 & 0.57 &  -24.7 \\
1997 Jul 16 & 43 & Core  & 10.21 &  ... &    ... & 0.05 & 1.00 &    ... \\ 
            &    & C9    &  2.62 & 0.22 & -143.7 & 0.00 & 1.00 &    ... \\ 
            &    & C8    &  4.09 & 0.36 & -133.3 & 0.04 & 1.00 &    ... \\ 
            &    & C7a   &  0.97 & 0.48 & -123.4 & 0.07 & 1.00 &    ... \\ 
            &    & C5    &  0.08 & 1.09 & -123.9 & 0.18 & 1.00 &    ... \\ 
            &    & C4    &  0.73 & 3.25 & -115.1 & 0.74 & 0.33 &  -14.1 \\ 
            &    & C4*   &  1.26 & 3.36 & -114.4 & 0.24 & 0.59 &  -12.3 \\
1997 Nov 16 & 43 & Core  & 21.06 &  ... &    ... & 0.05 & 1.00 &    ... \\ 
            &    & C9    &  3.39 & 0.17 & -145.7 & 0.04 & 1.00 &    ... \\ 
            &    & C8    &  4.19 & 0.38 & -137.2 & 0.08 & 1.00 &    ... \\ 
            &    & C7a   &  1.81 & 0.51 & -125.5 & 0.08 & 1.00 &    ... \\ 
            &    & C7    &  0.14 & 0.87 & -125.5 & 0.00 & 1.00 &    ... \\ 
            &    & C4    &  0.90 & 3.37 & -118.0 & 0.29 & 1.00 &    ... \\ 
            &    & C4*   &  1.31 & 3.40 & -113.5 & 0.18 & 1.00 &    ... \\ \tableline
\end{tabular}}
\end{center}
\tablenotetext{a}{Multiple IDs listed for a single Gaussian indicate a blending
of jet components.  Jet component C4 sometimes required multiple Gaussians
to fit.  The brightest Gaussian making up C4 is marked with an asterisk, and
its position is used as the position of C4.}
\tablenotetext{b}{Flux density in Janskys.}
\tablenotetext{c}{$r$ and PA are the polar coordinates of the
center of the Gaussian relative to the presumed core.
Position angle is measured from north through east.}
\tablenotetext{d}{$a$ and $b$ are the FWHM of the major and minor axes of the Gaussian.}
\tablenotetext{e}{Position angle of the major axis measured from north through east.}
\end{table*}

The Gaussian models are identified with named jet components in Table~\ref{mfittab}.
In all we identify a total of eight distinct jet components that we name C4, C5, C5a, C6, C7,
C7a, C8, and C9 from the outermost component inward.  A few words must be said about the naming
convention chosen for the jet components.  In all cases, we have tried to be consistent with naming
schemes adopted by other authors, and this has sometimes required the addition of a component name
with the ``a'' suffix in cases where a component was missed due to the sporadic single-epoch observations
by various authors.  Note also that there are two naming conventions in use for jet components
in 3C\,279.  The component named ``C5'' by C93 and Abraham \& Carrara (1998)
is not the same component referred to as ``C5'' by Wehrle et al. (1994),
Lepp\"{a}nen et al. (1995), and Lister et al. (1998).
We have not detected the ``C5'' component reported by C93 and Abraham \& Carrara (1998),
so in this paper we adopt the naming scheme of Wehrle et al. (1994),
Lepp\"{a}nen et al. (1995), and Lister et al. (1998).

We caution that Gaussian model fitting does not always produce unique results, especially for regions
of nearly continuous jet emission.
Blending of jet components into single Gaussians can be a problem at lower resolutions.
Component C5 is stationary at about 1 mas from the core, and as moving jet components approach
C5, the distinction between whether a given Gaussian model represents C5 or the approaching,
moving component becomes ambiguous.  Three closely spaced components in the inner jet (C7a, C8, and C9) are
detected in the 43 GHz data.  The 22 GHz model fits tend to represent these three jet components with
only two Gaussians, so further discussion of C7a, C8, and C9 is derived from the 43 GHz model fits.
The bright jet component C4 is detected at all epochs, and as it moves out its internal structure
becomes more complex, and more than one Gaussian is frequently required to represent it.
These multiple-Gaussian fits to C4 nearly always take the form of a brighter, more compact leading edge 
at a more northerly position angle, and
a fainter, more diffuse trailing Gaussian at a more southerly position angle.  
The position of the brightest Gaussian is used to represent the position of C4, these
Gaussians are marked with asterisks in Table~\ref{mfittab}.
The sharp leading edge of this component is suggestive of a working surface or shock front.  The
polarization observations of Lepp\"{a}nen et al. (1995), Lister et al. (1998), and Homan \& Wardle (1999)
show that C4 has a magnetic field transverse to the jet, also indicative of a shock front.

\subsection{Apparent Radial Motion}
\label{radmotion}

In this section we discuss the apparent radial motion of the jet components
identified in the previous section.  All of the jet components display apparent superluminal
motions with the exception of C5, which appears stationary at about 1.2 mas from
the core, and C9, which is present in the last 3 epochs at 43 GHz at about 0.2 mas
from the core.  The moving components interior to C5 tend to fade and disappear 
below our detection limit as they approach C5, although there are some indications
that they may remain barely detectable beyond this point (e.g. the emission beyond
C5 in the 1997 Nov 16 map at 22 GHz).  Three of the inner jet components (C5a, C6, and C7)
have undergone this behavior during our monitoring.

Motion of jet components in 3C\,279 can be seen directly from the images.
Figures~3 and 4 show a selection of images at 22 and 43 GHz respectively.
Only a selection of images are shown because there are too many
images to fit on a single page.
These images are presented as time-series mosaics with dotted lines indicating the best fits to
motion at constant speed for the model-fit Gaussians.  Animations made from the entire sequences
of images may be obtained from ftp://sgra.jpl.nasa.gov/pub/users/glenn/3c279\_22ghz.mpeg
and 3c279\_43ghz.mpeg.  

\begin{figure*}
\plotfiddle{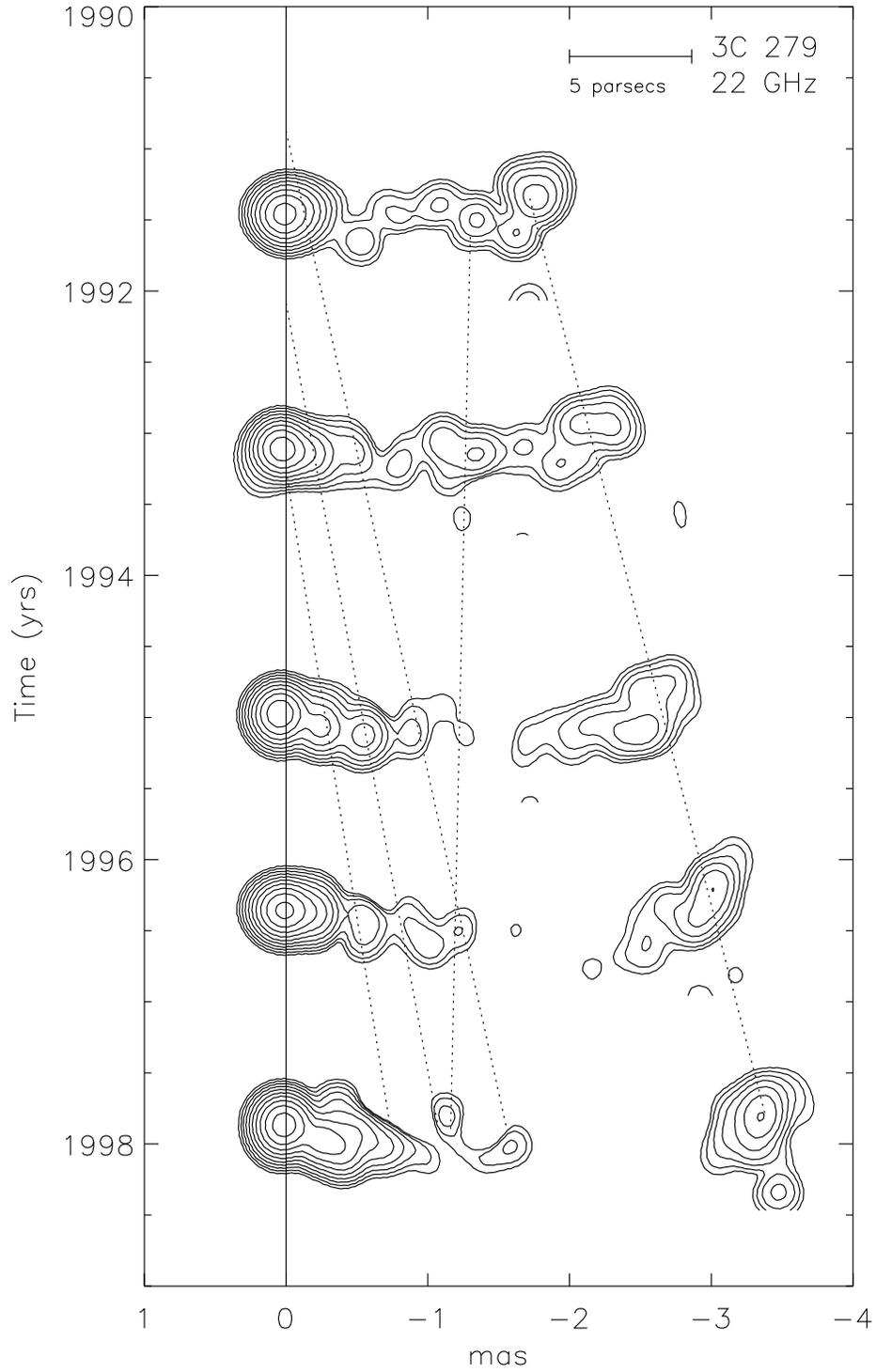}{8.0 in}{0}{75}{75}{-232}{0}
\caption{Time-series mosaic of a selection of 22 GHz VLBI images of 3C\,279.  Epochs
1991 Jun 24, 1993 Feb 17, 1994 Sep 21, 1996 May 13, and 1997 Nov 16 are shown.
The images have been restored with a circular 0.2 mas beam without residuals and
rotated 25$\arcdeg$ counterclockwise.  The lowest contour is 25 mJy beam$^{-1}$;
subsequent contours are a factor of two higher than the previous contour.
The solid line indicates the position of the presumed stationary core.  The dotted lines
represent the best fits to the model-fit Gaussian positions vs. time from Figure~5$a$.
From right to left these lines represent C4, C5, C5a, C6, and C7.  Some lines have been extended
before and after model-fit detections to show speculative zero-separation epochs and
later positions.}
\end{figure*}

\begin{figure*}
\plotfiddle{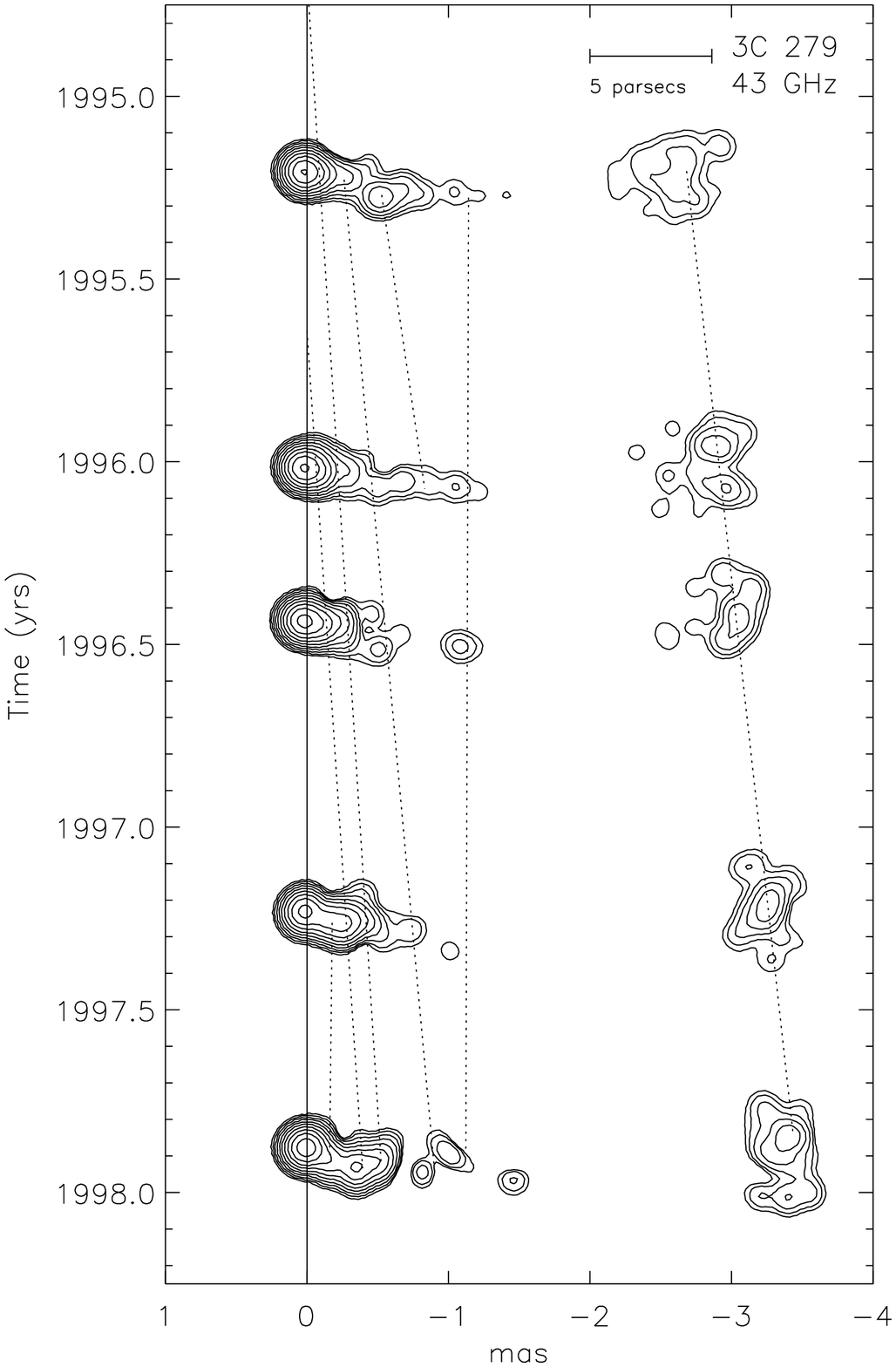}{8.0 in}{0}{75}{75}{-232}{0}
\caption{Time-series mosaic of a selection of 43 GHz VLBI images of 3C\,279.  Epochs 
1995 Mar 19, 1996 Jan 7, 1996 Jun 9, 1997 Mar 29, and 1997 Nov 16 are shown.
The images have been restored with a circular 0.15 mas beam without residuals and
rotated 25$\arcdeg$ counterclockwise.  The lowest contour is 25 mJy beam$^{-1}$;
subsequent contours are a factor of two higher than the previous contour.
The solid line indicates the position of the presumed stationary core.  The dotted lines
represent the best fits to the model-fit Gaussian positions vs. time from Figure~5$b$.
From right to left these lines represent C4, C5, C6, C7, C7a, C8, and C9.  Some lines have been extended
before model-fit detections to show speculative zero-separation epochs.}
\end{figure*}

The measured model-fit Gaussian
positions versus time are shown in Figures~5$a$ and $b$ for the 22 and
43 GHz model fits respectively.  
Error bars on the Gaussian positions have been set to be proportional to 
the beam size and inversely proportional to
the square root of the dynamic range of the component detection
(defined here as the uniform Gaussian surface brightness divided by the rms residual surface brightness),
and normalized to be one-quarter of a beam at a dynamic range of 200
by observing the scatter in the Gaussian positions.
Although the error may formally be inversely proportional to the dynamic range (rather than the square root)
(e.g. Fomalont 1985; Reid \& Moran 1988; Condon 1997), the range in component surface brightnesses
in 3C\,279 is so large (of order 1000 between C8 and C5) that this produces an extremely large
range in errors that does not empirically agree with the scatter in the model-fit positions.
Effects other than visibility errors (e.g., $(u,v)$ coverage effects, 
calibration errors, and interactions among multiple model components) presumably dominate
at the higher surface brightnesses and simulate a dependence with lower slope, and we find
that the square root of the dynamic range empirically provides a good estimate for the errors.
Some of the scatter about straight-line fits may also be real and due to intrinsic or
viewing angle produced accelerations or changes in the complex underlying internal brightness
distribution of the convolved components (G\'{o}mez et al. 1997).

\begin{figure*}
\plotfiddle{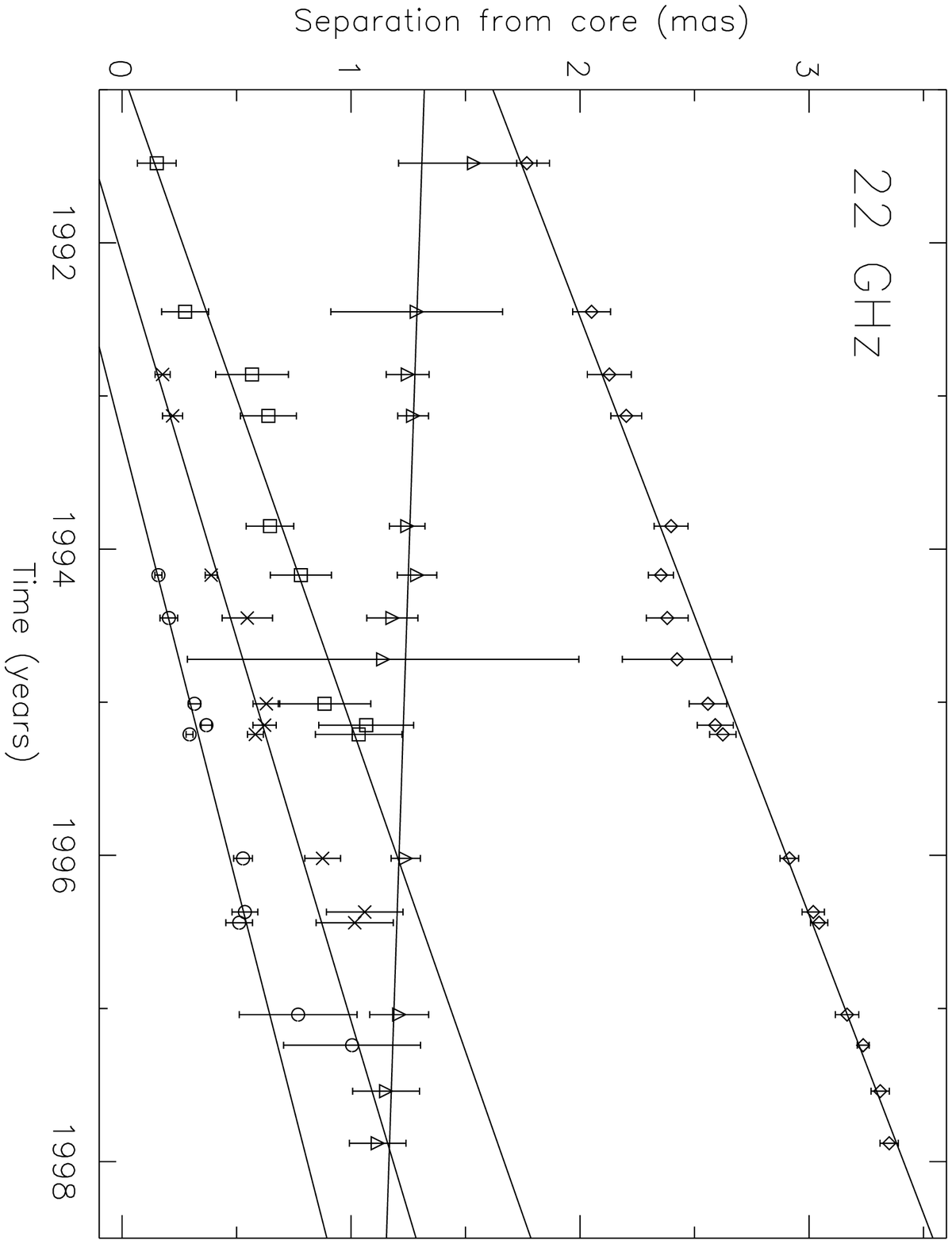}{3.375 in}{90}{60}{60}{227}{-54}
\end{figure*}

\begin{figure*}
\plotfiddle{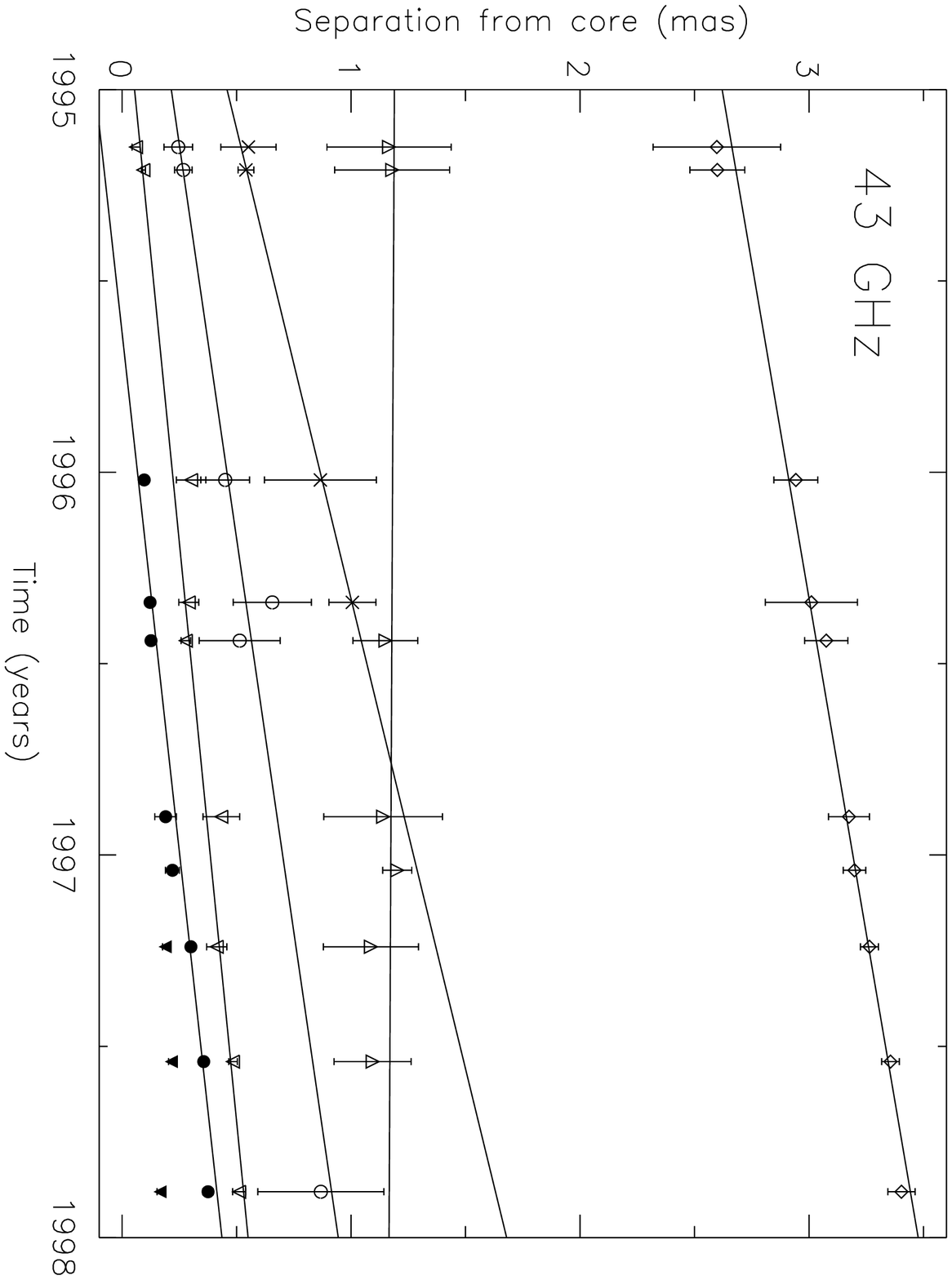}{3.375 in}{90}{60}{60}{227}{-27}
\caption{Distances from the core of model-fit Gaussians in 3C\,279 as a function of time.
Diamonds represent component C4, upward-pointing triangles C5, squares C5a, crosses C6,
circles C7, downward-pointing triangles C7a, filled circles C8, and filled
downward-pointing triangles C9.  The lines shown are the best fits to motion with
constant speed. ($a$) 22 GHz. ($b$) 43 GHz.}
\end{figure*}

Changes in the jet emission from epoch to epoch can also be examined using flux
profile curves (summed flux perpendicular to the jet).  Figures~6$a$ and $b$ show
the flux profile at each epoch plotted as a variable-width black bar, with
diagonal lines indicating fits to
model-fit Gaussian positions versus time.  Motions of well-separated components like
C4 are easy to see in these plots, while motions of inner components which often appear as
extensions or ``shoulders'' on a declining jet are more difficult to see.

\begin{figure*}
\plotfiddle{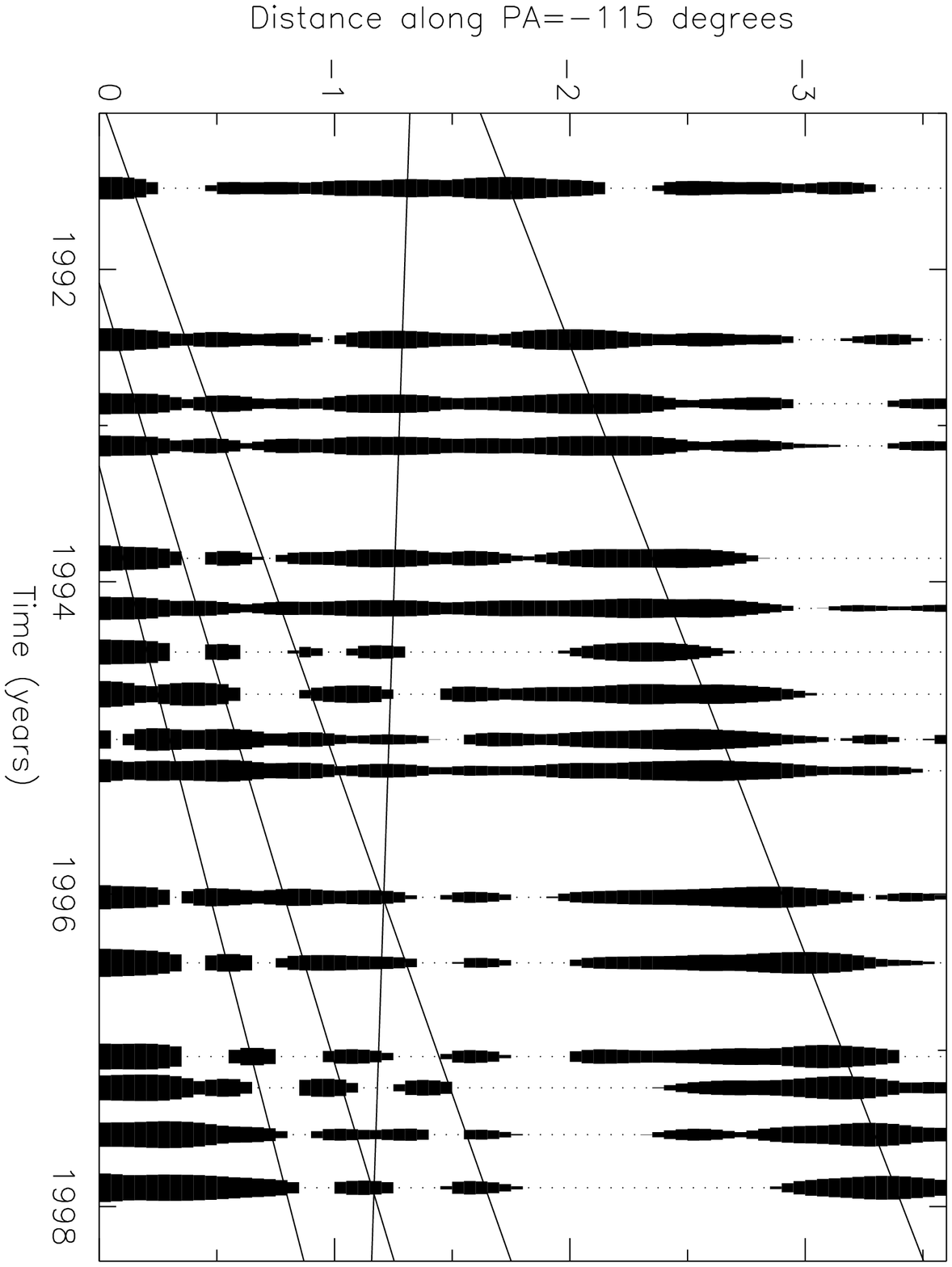}{3.375 in}{90}{60}{60}{227}{-54}
\end{figure*}

\begin{figure*}
\plotfiddle{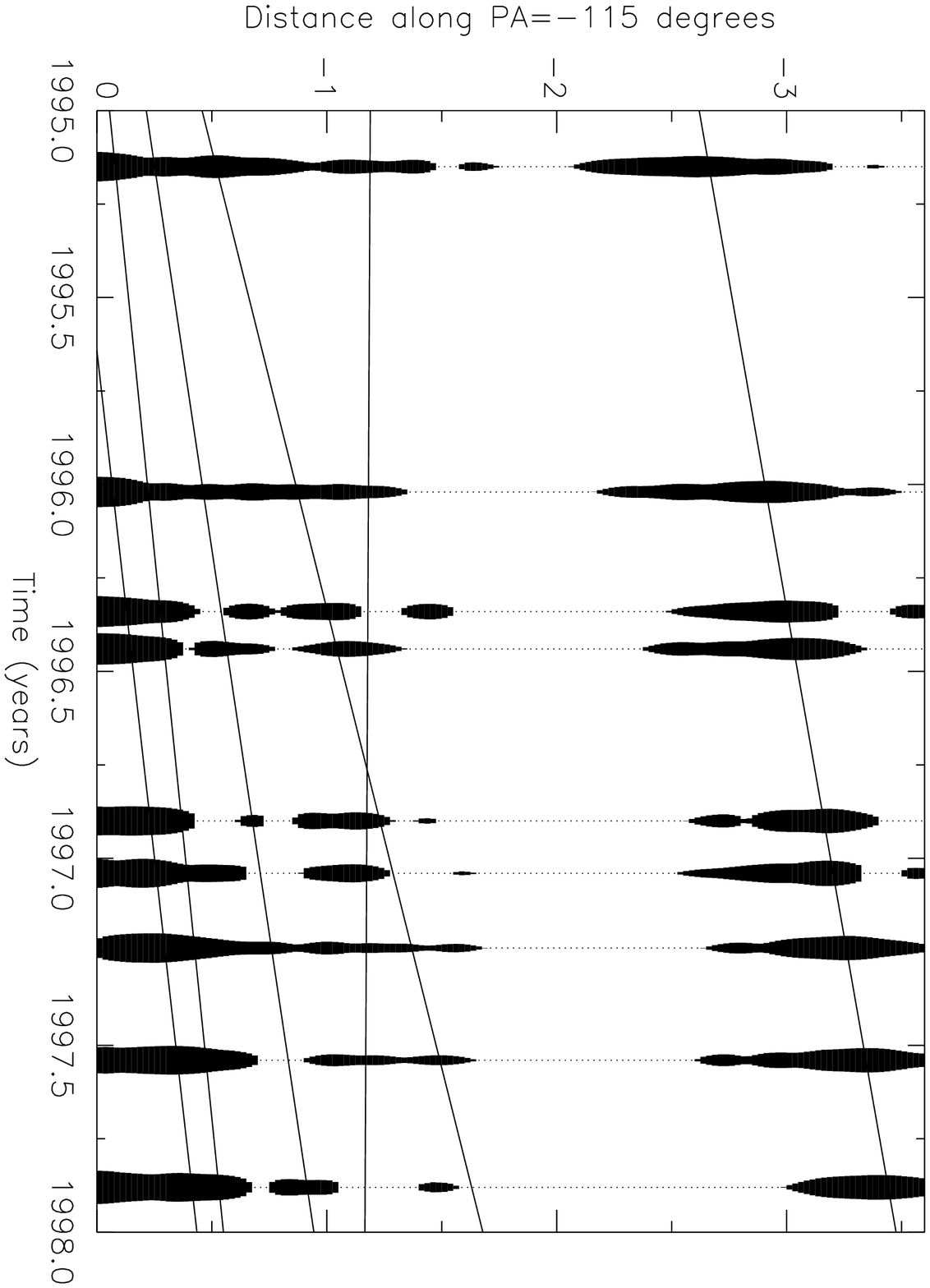}{3.375 in}{90}{60}{60}{227}{-27}
\caption{Flux profiles of the 3C\,279 jet vs. time.
The width of the variable-width black bar plotted for each epoch is proportional
to the logarithm of the flux summed perpendicular to the jet (assumed to be at a position
angle of $-115\arcdeg$), at a distance along the jet indicated by the $y$-axis.
A 5 Jy delta function at zero mas (representing the core) and a smoothly declining exponential jet have been
subtracted from the fluxes to enhance the contrast of jet components.
The solid lines show the best fits to the model-fit Gaussian positions vs. time from Figures~5$a$
and $b$.  ($a$) 22 GHz.  All epochs are shown with the exceptions of 1995 Feb 25 and
1996 May 13.  ($b$) 43 GHz.  All epochs are shown with the exception of 1995 Mar~19.}
\end{figure*}

Apparent speeds of the components were measured by performing least-squares fits to the Gaussian 
radial positions versus time from Figures~5$a$ and $b$, 
assuming radially outward motion at constant speed.
The data indicate this to be a good approximation for all components except C4 
(see $\S$~\ref{3dc4}).  For four components (C4, C5, C6, and C7) we have measured speeds
at both 22 and 43 GHz.  The speeds at 22 and 43 GHz are consistent
(as they should be) for all of these components except C6, where the speeds at
22 and 43 GHz differ at the 2$\sigma$ level.  We believe this to be due to a biasing outward
of the position of C6 at 43 GHz during its short detection period at this frequency due to a
blending with C5, which it was then approaching.  For those components detected at both frequencies,
the most accurate speeds were obtained from the 22 GHz monitoring, due to its three-times longer
time baseline.  The measured speeds for all components are listed in Table~\ref{speedtable}.
These speeds differ significantly from each other, with the moving components
(excluding C5 and C9) becoming progressively slower from C4 to C7a, and the
speed increasing slightly again with C8.

\begin{table*}[!t]
\caption{Apparent Speeds\tablenotemark{a}}
\label{speedtable}
\begin{center}
\begin{tabular}{c r c c} \tableline \tableline
& \multicolumn{1}{c}{Speed} & Zero Separation & Frequency\tablenotemark{b} \\ 
Component & \multicolumn{1}{c}{($c$)} & Epoch & (GHz) \\ \tableline \\ [-5pt]
C4  & 7.5$\pm$0.2  & 1984.68$\pm^{0.27}_{0.29}$ & 22 \\ [3pt] 
C5  & $-0.6\pm$0.5 & ...                        & 22 \\ [3pt]
C5a & 6.8$\pm$1.0  & 1990.88$\pm^{0.28}_{0.39}$ & 22 \\ [3pt]
C6  & 5.8$\pm$0.4  & 1992.09$\pm^{0.14}_{0.17}$ & 22 \\ [3pt]
C7  & 5.0$\pm$0.4  & 1993.26$\pm^{0.13}_{0.14}$ & 22 \\ [3pt]
C7a & 4.8$\pm$0.2  & 1994.67$\pm^{0.04}_{0.05}$ & 43 \\ [3pt]
C8  & 5.4$\pm$0.2  & 1995.63$\pm^{0.07}_{0.07}$ & 43 \\ [3pt]
C9  & $-0.9\pm$0.8 & ...                        & 43 \\ [3pt] \tableline
\end{tabular}
\end{center}
\tablenotetext{a}{for $H_{0}$=70 km s$^{-1}$ Mpc$^{-1}$, $q_{0}$=0.1.}
\tablenotetext{b}{Frequency at which the speed measurement was made.}
\end{table*}

These speeds can be compared with those measured previously by other authors.
3C\,279 was one of the first quasars to have a superluminal speed measured.
Older VLBI observations (Cotton et al. 1979; U89; C93) followed a series of
components (C1--C3) moving along position angles of $-130\arcdeg$ to $-140\arcdeg$.
Cotton et al. (1979) reported a large proper motion of
0.5~mas yr$^{-1}$ (15$c$) during the early 1970s;
their data was re-examined and their interpretation confirmed
by U89.  U89 and C93
observed the motions of several new superluminal components, and
found that the speeds of these components were only one-quarter to one-third (3-5$c$)
of that measured for the original superluminal component during the 1970s.
Component C4 discussed in this paper was first detected in these observations
of U89 and C93 at its present position angle of $-114\arcdeg$.  
C93 determined a motion of 0.15$\pm$0.01~mas yr$^{-1}$ (4$c$) for C4,
significantly different from the speed measured in this paper.
There are a number of potential causes for this discrepancy:
at the time of those observations C4 was about 1 mas from the core
(near C5) in a region that is difficult to interpret, the VLBI data from
the late 1980s are of inferior quality to modern VLBA data (see Appendix A), and
the motion of C4 is inherently complex (see $\S$~\ref{3dc4}).  The speed measured
here for C4 agrees well with that measured in 15 GHz VLBA monitoring (Kellermann 1999, private communication).
The different speeds measured for various components in our monitoring
are consistent with the varying speeds observed for different components during the 1970s and 80s
and show that this variation in speed from component to component is a real property of this source.

\subsection{C5 and Stationary Components}
\label{stationary}
Component C5 has a fitted motion consistent with zero speed ($-0.6\pm0.5c$) over
the six years from 1991 to 1997 at 22 GHz, although it does show some evolution,
dimming considerably in flux over this time (see Figure 3).
Stationary jet components are a relatively common phenomenon in superluminal sources.
The well-known source 4C\,39.25 has long displayed a bright stationary component
in addition to the core (Alberdi et al. 1999).  Core-jet sources with stationary
components are sufficiently common that a sizable percentage of sources originally classified
as compact doubles were later found to be core-jet sources 
with stationary components rather than the physically
distinct compact symmetric objects (Conway et al. 1994).
Stationary components were also found to be a common feature of EGRET blazars
in the monitoring program of Marscher et al. (2000).

Stationary components could be due to two causes: a standing shock in the jet,
or a point along a curving jet where the jet crosses the line-of-sight.
In the second case the stationary component should appear quite bright, since the
jet material would have maximal Doppler boosting when passing this point.
Since C5 is at times quite faint, and since there appears to be a physical change
in the jet at about that point (with most superluminal components fading and disappearing
at about the location of C5 --- possibly becoming disrupted), we favor the explanation
of C5 as a standing shock in the jet.  G\'{o}mez et al. (1995) find that standing oblique
shocks are created in their numerical simulations of relativistic jets
by quasi-periodic recollimation shocks.  A standing shock may also be created through
a sudden variation in the external pressure that results in an isolated
recollimation shock, and it will be an interesting future study to see if 
isolated stationary components correspond to a physical point in the source where
the pressure would be expected to change abruptly.

The bright flux of C5 around 1991 and its subsequent decay may be due to
the lingering effects of its previous encounter with C4, which passed the location
of C5 in the late 1980s.  If the slight inward motion measured for C5 is real,
this may be due to a temporary dragging downstream of the stationary component
during its interaction with C4.  G\'{o}mez et al. (1997) find that both of these
effects (brightening and dragging downstream of the stationary component) occur when superluminal components
interact with stationary components in their relativistic jet simulations.

\subsection{Component Position Angles}
The kiloparsec-scale VLA jet of 3C\,279 mapped by de Pater \& Perley (1983) displays
several components with position angles ranging from $-$145\arcdeg to $-155\arcdeg$.
Position angles of the older VLBI components C1--C3 measured by 
Cotton et al. (1979), U89, and C93 were
similar: $-130\arcdeg$ to $-140\arcdeg$, implying slight bending between the parsec and
kiloparsec scales.  More recent high-dynamic range VLBA maps at lower frequencies
(5 and 1.6 GHz)  by Piner et al. (2000a) have confirmed the larger-scale VLBI jet
to have components with position angles ranging from $-135\arcdeg$ to $-155\arcdeg$, with
several small bends present between the core and the last visible component at
$\sim$100 mas.  The current dominant feature in the smaller-scale VLBI jet (C4)
has a position angle of $-114\arcdeg$, quite different from any of these larger-scale
position angles.  C4 shows no signs of altering its path
to follow the larger-scale structure, and appears to be continuing along a position angle of $-114\arcdeg$.
In fact, both the higher dynamic range 22 GHz images from this paper (see $\S$~\ref{sec-imaging})
and the lower-frequency VSOP and VLBA images from Piner et al. (2000a) show steep-spectrum diffuse emission
at a position angle of about $-140\arcdeg$ between the core and C4, indicating that C4 has
already passed the point where it would have needed to turn to follow the larger-scale
VLBI structure.

Figure~7 shows the position angles of C4 and the other inner jet components 
discussed in this paper.  Since no individual component shows a significant change in position angle
(with the exception of C4, where the change is smaller
than the scale important for this plot), we plot the average position angle for each component.
Figure~7 shows that the average position angles of jet components 
have changed during our monitoring.
The position angles of the new components C8 and C9 are similar
to that of the larger-scale VLBI structure.  Since these new components have only been 
monitored over a 1 to 2 year time frame, it remains to be seen if their position angles will
change with time to match any of the earlier components.  

\begin{figure*}[!t]
\plotfiddle{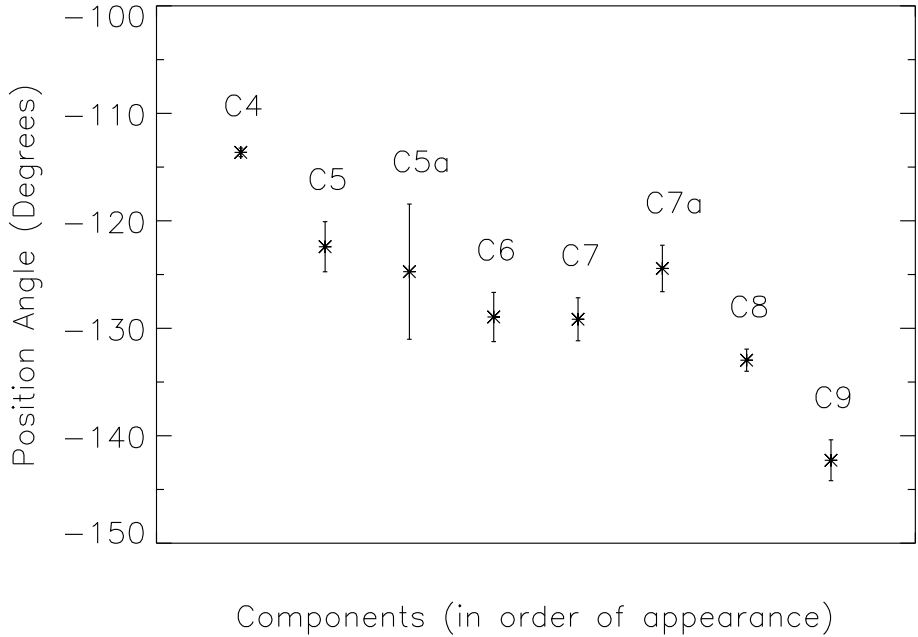}{3.5 in}{0}{75}{75}{-223}{-272}
\caption{Average position angle for each jet component discussed in this paper,
in order of appearance.}
\end{figure*}

Abraham \& Carrara (1998) attempted to explain the different apparent speeds and position angles
of the older VLBI components by postulating a precessing nozzle that ejects components at
varying position angles and viewing angles.  While we confirm the presence of different position
angles and apparent speeds for different components, our measurements do not agree
with an extrapolation of their specific model to the epochs observed in this paper.  Our measured apparent
speeds agree reasonably well with an extrapolation of their Figure~2 (they predict a slowing of 
apparent speeds for ejections between 1989 and 1996), but the change in component position angles
from Figure~7 does not agree with their Figure~1, and even has an opposite slope.
Since the different apparent speeds and position angles are suggestive of a precessing nozzle,
it may be possible for a different precession model to fit the current data; however,
if component ejections really vary regularly between $-114\arcdeg$ and $-142\arcdeg$ on $\sim$10 year
timescales as indicated by Figure~7, it is difficult to understand why the larger-scale
VLBI and VLA emission should be constrained to the narrow range of position angles observed.

\subsection{3-Dimensional Trajectory of Component C4}
\label{3dc4}
We have a large number of precise measurements of component C4
(eighteen at 22 GHz and ten at 43 GHz), which enable
us to track its motion in detail.  We are fortunate that this component
has been unusually bright and long-lived, and has been visible over the entire
six years of monitoring at 22 GHz.  Component C4 has followed
a curved path over the years 1991 to 1997 (an indication of this can be seen in Figure~5$a$,
where the apparent speed appears to slow around 1994). 
We use this curved path and measured apparent speeds to derive the complete 3-dimensional
trajectory of C4,
using the method described by Zensus, Cohen, \& Unwin (1995) for the motion of
components in 3C\,345.  This method is displayed graphically in Figure~8.
We use the positions measured in the 22 GHz monitoring, since this has
a time baseline three times longer than the 43 GHz monitoring.

\begin{figure*}
\plotfiddle{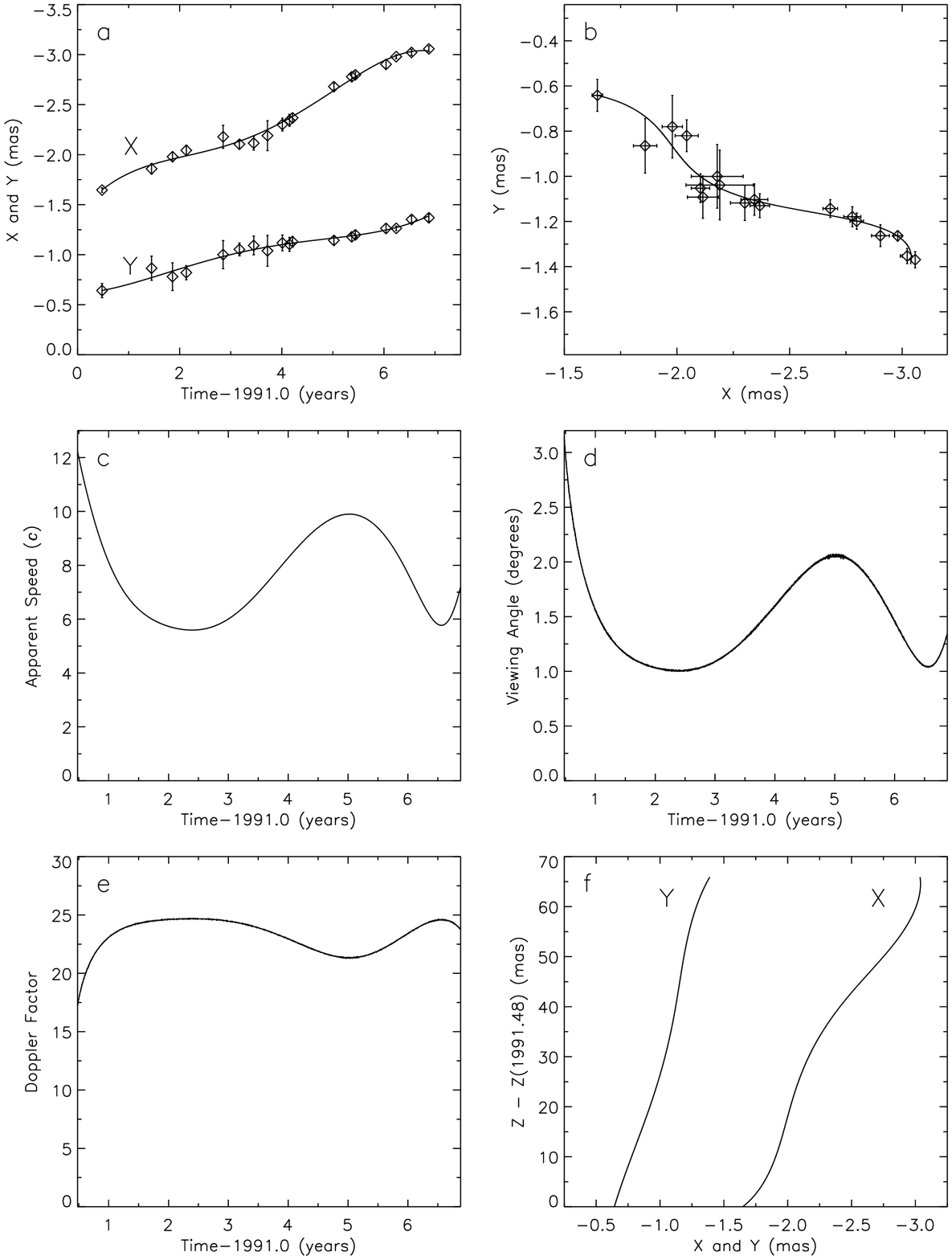}{8.0 in}{0}{80}{80}{-248}{0}
\caption{Derivation of the complete 3-dimensional trajectory of component C4 at 22 GHz
from 1991 to 1997.  Individual panels are discussed in the text of $\S$~\ref{3dc4}.}
\end{figure*}

We start by deriving weighted least-squares polynomial fits to $x(t)$ and $y(t)$,
the measured offsets from the core in the plane of the sky at epoch $t$.  
The measured $x(t)$ and $y(t)$ values and their associated fourth-order fits are shown in Figure~8$a$.
A straight line is a very bad fit to the measured $x(t)$ values 
(reduced chi-squared = 2.8), and a fourth-order polynomial is the lowest order
that achieves a good fit (reduced chi-squared = 0.71).  Zensus et al. (1995) also
needed fourth-order polynomials to fit the motion of components in 3C\,345.
Figure~8$b$ shows the same measured values and fits from Figure~8$a$ in the $(x,y)$ plane,
and shows the substantial curvature in the C4 trajectory.

From the fits to  $x(t)$ and $y(t)$ we derived the instantaneous apparent speed 
$\beta_{app}(t)\propto(\dot{x}(t)^{2}+\dot{y}(t)^{2})^{1/2}$, which is shown in Figure~8$c$.
The apparent speed varies between 6 and 12$c$ over the years 1991 to 1997.  Note that
Figure~8$c$ shows the magnitude of the total velocity vector (this is more appropriate
for a component following a curved trajectory), unlike Figure~5$a$ which shows only the 
radial component.  These changes in $\beta_{app}$ with time could be caused either by
changes in the Lorentz factor $\Gamma$ or the viewing angle $\theta$ (or both).  Since we know the
trajectory to be undergoing bends in the sky plane, a constant $\theta$ would require
a special alignment of the true bends with respect to the observer, which we consider unlikely.
We therefore take the other simplifying assumption (constant $\Gamma$), and assume that
the changes in $\beta_{app}$ with time are due to changes in $\theta$.  We assume $\Gamma$
to be constant at a value slightly larger than the minimum $\Gamma$ required for
the maximum $\beta_{app}(t)$: $\Gamma_{min}=(\beta_{app}(t)^{2}+1)^{1/2}$,
so we take $\Gamma=13$.  Under the assumption
of constant $\Gamma$, we can use the observed $\beta_{app}(t)$ to calculate $\theta(t)$.
Since the equation for $\theta(t)$ is quadratic, we must
choose either the large or the small-angle solution, this is shown in Figure~8$d$.  
We take the small-angle solution because the large-angle
solution gives large changes in the Doppler factor $\delta$ that would be seen as large changes in the flux
of C4, and these are not observed.  Figure~8$e$ confirms that $\delta(t)$ shows little variation
in the case of the small-angle solution.

The small-angle solution constrains C4 to viewing angles of $1\arcdeg$ to $3\arcdeg$.
Small viewing angles such as these are typical of those expected for an EGRET blazar
(see the Monte Carlo simulations of Lister [1998]), and angles in this range have been
successfully used to model the multiwavelength spectrum of 3C\,279 by, e.g., 
Maraschi, Ghisellini, \& Celotti (1992) ($3\arcdeg$), Ghisellini \& Madau (1996) ($4\arcdeg$),
and Hartman \& Boettcher (1999) ($2\arcdeg$).  Once $\theta(t)$ is known the trajectory
of the component in the $(x,z)$ and $(y,z)$ planes (where the $z$-axis
is perpendicular to the plane of the sky) can be calculated.  These trajectories are shown
in Figure~8$f$ (note the expanded scale along the $z$-axis), 
together with Figure~8$b$, they describe the complete 3-dimensional motion
of the component.  Component C4 appears to describe approximately
one oscillation of a low-pitch angle helical motion, notable especially in the
$(x,z)$ plane trajectory.  Helical motion of components may be caused
by hydrodynamic instabilities (Hardee 1987) or magneto-hydrodynamic effects 
(Camenzind 1986; Camenzind \& Krockenberger 1992), and has previously been observed
in several sources, notably 3C\,345 (Steffen et al. 1995) and BL Lac
(Tateyama et al. 1998; Denn \& Mutel 1999).

The bending of the trajectory of C4 is much smaller (about 4$\arcdeg$ total change
in position angle) than the position angle differences between components shown in Figure~7.
This indicates that a complete description of component motions in 3C\,279 may require a superposition of small-scale
wiggles --- possibly due to hydrodynamic instabilities --- such as those observed in the motion
of C4, and a larger-scale motion of the jet nozzle that causes the differences in 
apparent speed and position angle from component to component.  This would make a 
complete description of motions in 3C\,279 quite complex, and is reminiscent of the complexity
observed in 3C\,345, where different components follow different curved paths
(Zensus et al. 1995).  Detailed motions of components in 3C\,279 other than C4
are difficult to observe, since other components tend to fade and disappear within about 1 mas of
the core.  To further constrain the dynamics of the 3C\,279 jet it will be necessary to continue monitoring
the position of component C4 as it evolves, the bright component C8 to see if it continues
past the 1 mas point (which it will reach in about 2002), and the position angles of new
ejections.

\subsection{Jet Opening Angle}
We can measure the apparent opening angle of the 3C\,279 jet using the model-fit
sizes of the core and component C4.  Component C4 is
well-resolved transverse to the jet (see Figures~3 and 4); at many epochs an elliptical Gaussian 
with major axis transverse to the jet is required to
model it.  We used the mean core size at 22 GHz --- 0.07 mas, which is partially resolved, see
$\S$~\ref{tbmeassec} --- and calculated the apparent opening angle of C4 for each 22 GHz epoch.
The median apparent full-cone opening angle is 12$\arcdeg$.  This angle can be deprojected using
a typical viewing angle for C4 of 2$\arcdeg$ (an approximate upper limit since viewing 
angle was derived using a minimum $\Gamma$), estimated from Figure~8$d$, to obtain an 
intrinsic full-cone opening angle of 0.4$\arcdeg$.  This opening angle is quite small; for comparison,
Unwin et al. (1994) derived a full-cone opening angle for 3C\,345 of 5$\arcdeg$, although
this number was not well constrained.
Marscher et al. (2000) mention that a number of the jets of EGRET blazars are quite
broad within a few mas of the core, which suggests that they are being viewed nearly end-on.
The inner jet of 3C\,279; however, is relatively narrow (see Figures~3 and 4), despite
its status as an EGRET blazar, confirming an intrinsically small opening angle as derived above.
G\'{o}mez et al. (1997) find small full-cone opening angles of 0.8$\arcdeg$ in their numerical
simulations of both pressure-matched and overpressured relativistic jets.

\subsection{Zero-Separation Epochs and EGRET Light Curve}
In this section we examine whether any relationships can be
demonstrated between the ejection epochs of the superluminal
components and $\gamma$-ray high states or ``flares'' observed by
CGRO EGRET.  Such a relationship could indicate that the $\gamma$-ray high
states correspond to major disturbances in the energy flow of the
relativistic jet.  In the VLBA EGRET-blazar monitoring campaign
described by Marchenko-Jorstad et al. (2000), correlations between
VLBI component ejections and $\gamma$-ray flares are seen in 10 out of
19 cases of $\gamma$-ray flares for which sufficient VLBI data existed
to detect such a correlation, a number far exceeding that expected by
random chance.

We have extrapolated the epochs of zero-separation from the core
(i.e., centers coincident) of the five moving components that have
emerged during the EGRET era (C5a, C6, C7, C7a, and C8) by assuming
motion at the constant radial speeds given in $\S$~\ref{radmotion}.
The estimated epochs of zero-separation are listed in
Table~\ref{speedtable}.  Errors on the zero-separation epochs were
derived from the errors in the speeds quoted in
Table~\ref{speedtable}. The component zero-separation epochs 
can be used to calculate the intervals between component
ejections.  For the five components listed above (C5a, C6, C7, C7a,
and C8) the four corresponding ejection intervals are 1.21$\pm$0.33,
1.17$\pm$0.20, 1.41$\pm$0.14, and 0.96$\pm$0.08 years.  While not
statistically identical, these times do suggest an ejection rate of
about one component per year.

These epochs of zero-separation are compared with the EGRET ``light
curve'' in Figure~9.  The EGRET fluxes have been taken from Hartman et
al. (1999) (prior to 1996), and Hartman (2000, private communication)
(1996 and later).  Recall that EGRET viewing periods varied in length
from about two weeks to seven weeks; in some periods, the full viewing
period was needed to collect enough data for a single reliable
detection or to form an upper limit.  ``Flare'' indicates a high
state, not that a well-defined rise and decline from a quiescent level
was observed.  Figure~9 clearly shows that any statistical study is
hampered by the sparseness of the EGRET light curve.  Inspection of
Figure~9 shows that there is no overlap between any of the 1$\sigma$
ranges on component zero-separation epochs and an EGRET viewing period
on 3C\,279. In other words,  we were never observing with EGRET
when a VLBI component was being ejected.

\begin{figure*}[!t]
\plotfiddle{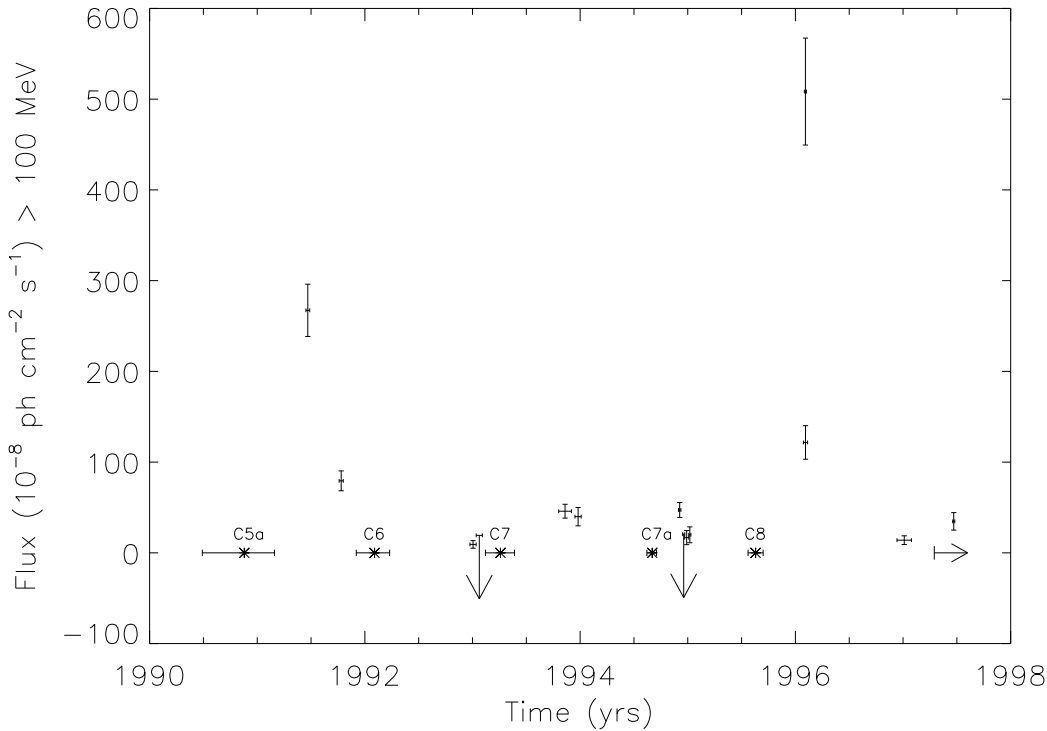}{3.875 in}{90}{60}{60}{227}{-40}
\caption{EGRET light curve compared to VLBI component zero-separation epochs.
EGRET fluxes were taken from Hartman et al. (1999) and Hartman (2000, private communication).
Upper limits shown are 2$\sigma$.  A residual uncertainty of 10\% was used in the flux errors
in addition to the statistical uncertainty as suggested by Hartman et al. (1999).
Combined viewing period fluxes have been used
where available and where the individual viewing period fluxes are consistent
with a constant flux over the summed interval.
Estimated zero-separation epochs of the VLBI components assuming constant speed
are plotted as asterisks with accompanying error bars along the zero-flux line.
The arrow near the right edge of the plot indicates the last epoch at which a component
ejection could have been detected, assuming a component speed of 5$c$ and detection at 0.1 mas.}
\end{figure*}

There {\it are} components ejected close to the times of the major
$\gamma$-ray ``flares'' in 1991 and 1996, but this may have occurred by
random chance.  Component C5a's zero-separation epoch is
1990.88$\pm^{0.28}_{0.39}$, prior to the $\gamma$-ray high state
observation at 1991.47.  Component C8's zero-separation epoch is
1995.63$\pm^{0.07}_{0.07}$, prior to the $\gamma$-ray high state
observation at 1996.09.  A correlation cannot be demonstrated
statistically because the EGRET data are too sparsely sampled,
however, ``Absence of evidence is not evidence of absence.''  The
next-generation $\gamma-$ray satellite, GLAST, will be much more suited
to statistical studies.

\section{FLUX AND SPECTRAL EVOLUTION OF THE CORE AND JET}
\label{sec-radioevolution}

\subsection{Radio Light Curves}
\label{lightcurves}
In this section we discuss the relation between the evolution in brightness
of individual VLBI components and that of the total source flux.
The motivation for this approach is to take
advantage of the very long history of flux monitoring of 3C\,279.  Regular
monitoring at centimeter wavelengths was begun
by the Michigan group in 1965 (Aller et al. 1985), and
the source is now monitored at intervals of a few days to frequencies as
high as 87 GHz.

It is clear from the Michigan radio light curves at 4.8, 8.0
and 14.5 GHz that the character of the flux curves changed in
1990.  Prior to 1990, individual flux outbursts were well separated, by
intervals of up to 5 years.  The 1981-1985 burst was well fitted by a shock
model in which the low-frequency evolution lagged the higher frequencies
(Hughes et al. 1991).  Until 1990, the spectrum between 8 and 14.5 GHz
remained essentially flat ($\alpha \sim 0$), but since then, the spectrum has
been strongly inverted.  Even more striking, the characteristic timescale
for flux outbursts has decreased to less than a year, with detectable changes
on scales of a week or less.

This increase in complexity of the total flux light curves does not
correspond to any increase in the complexity of the VLBI images.
The simplest explanation
is that the bulk of the flux variation occurs in the almost-unresolved core.
Figure 10$a$ shows the flux density evolution of the core
and jet components from model fitting at 22 GHz.  Also shown are the 22 GHz total flux
density monitoring data from Mets\"{a}hovi.  Figure 10$b$ shows the same data over
a smaller range in flux so that the variability of the jet components can be seen.
Similarly, Figure 10$c$ shows the
43 GHz core and jet component fluxes and 37 GHz Mets\"{a}hovi data, and
Figure 10$d$ shows the same data over a smaller range in flux.
We estimate the amplitude calibration of these VLBI fluxes to be accurate
at about the 10\% level.

\begin{figure*}
\plotfiddle{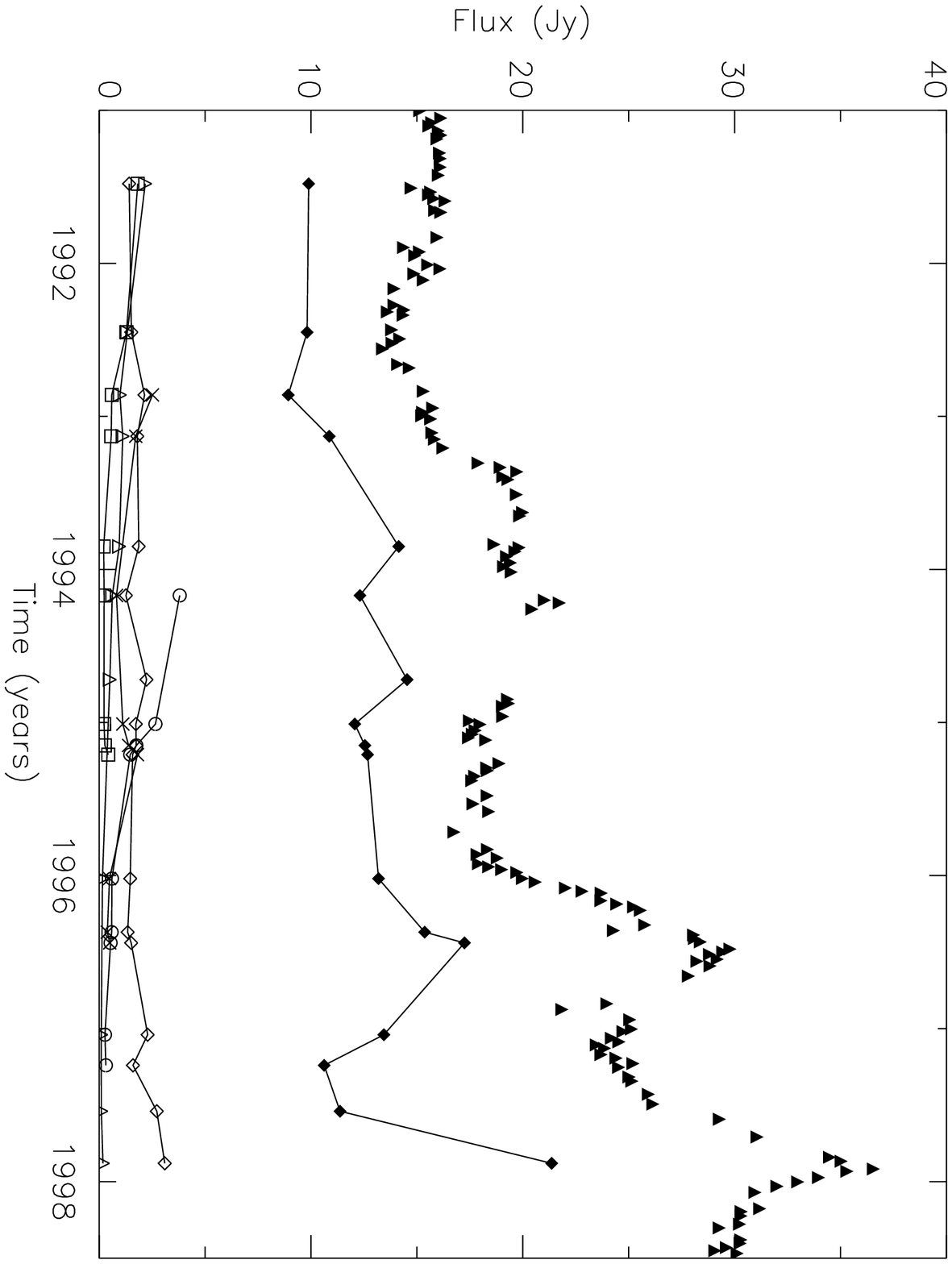}{4.2 in}{90}{60}{60}{227}{-18}
\end{figure*}

\begin{figure*}
\plotfiddle{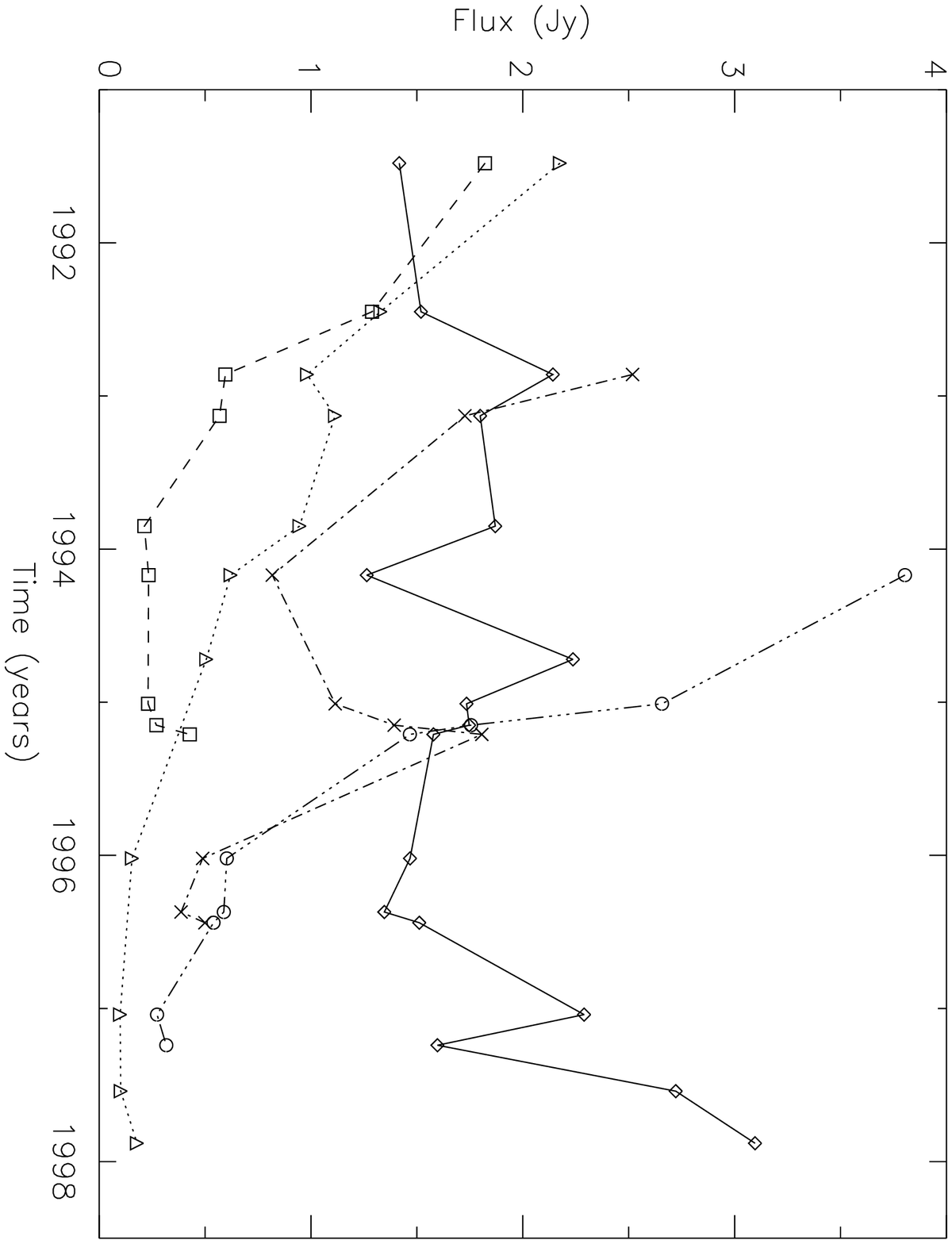}{3.375 in}{90}{60}{60}{227}{-27}
\caption{VLBI and total flux light curves of 3C\,279.  The filled upward-pointing triangles in
Figures $a$ and $c$
represent total flux density monitoring data from Mets\"{a}hovi at 22 and 37 GHz respectively.
Other symbols represent VLBI model-fit fluxes: filled diamonds the core,
open diamonds C4, upward-pointing triangles C5, squares C5a, crosses C6,
circles C7, downward-pointing triangles C7a, filled circles C8, and filled
downward-pointing triangles C9.  The poor sensitivity observation on 1994 Jun 12
is not included.
($a$) 22 GHz.  ($b$) 22 GHz with an expanded flux scale to
show variability of jet components. ($c$) 43 GHz.  ($d$) 43 GHz with an expanded flux scale to
show variability of jet components.}
\end{figure*}

\begin{figure*}
\plotfiddle{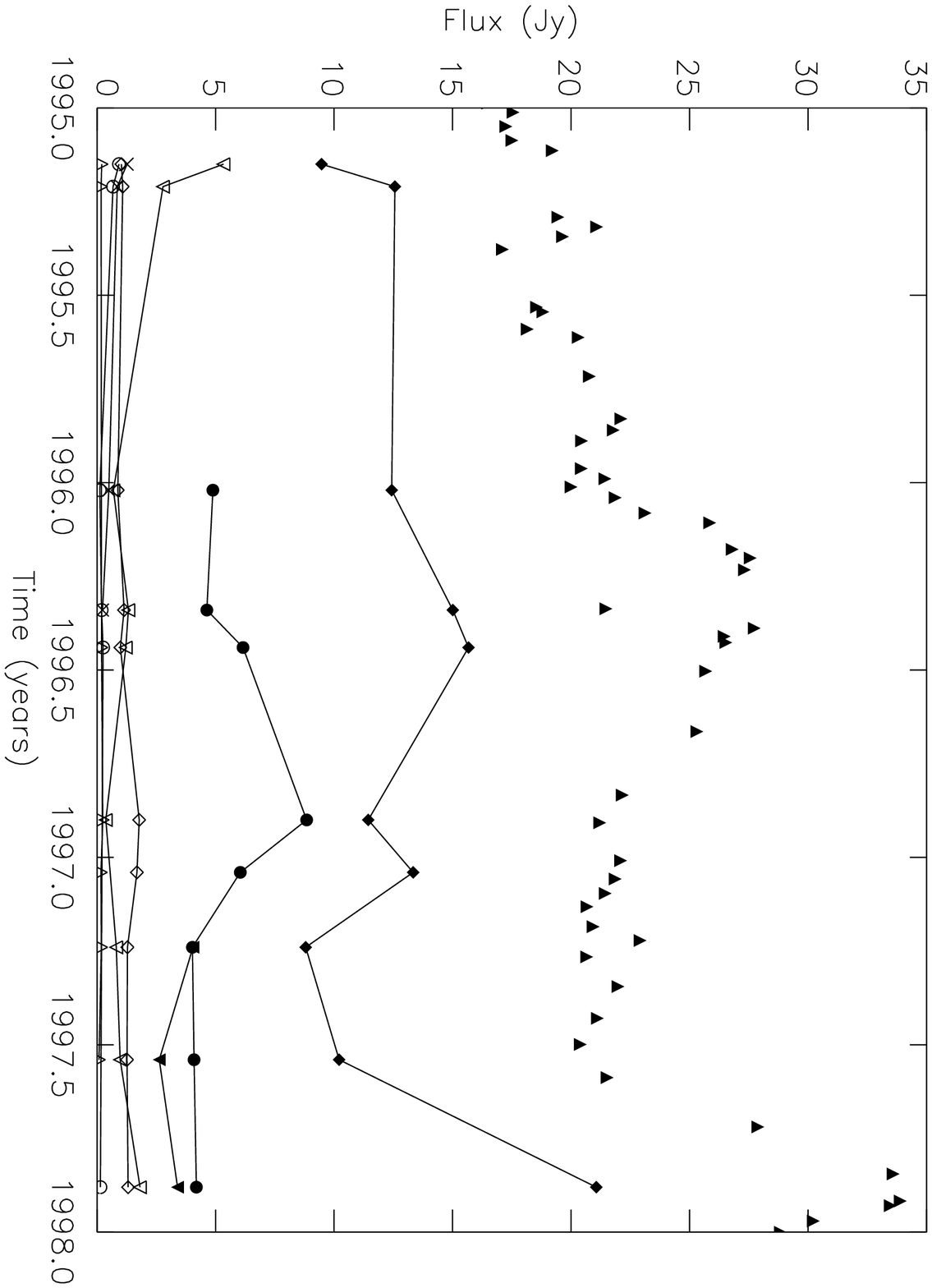}{3.375 in}{90}{60}{60}{227}{-54}
\end{figure*}

\begin{figure*}
\plotfiddle{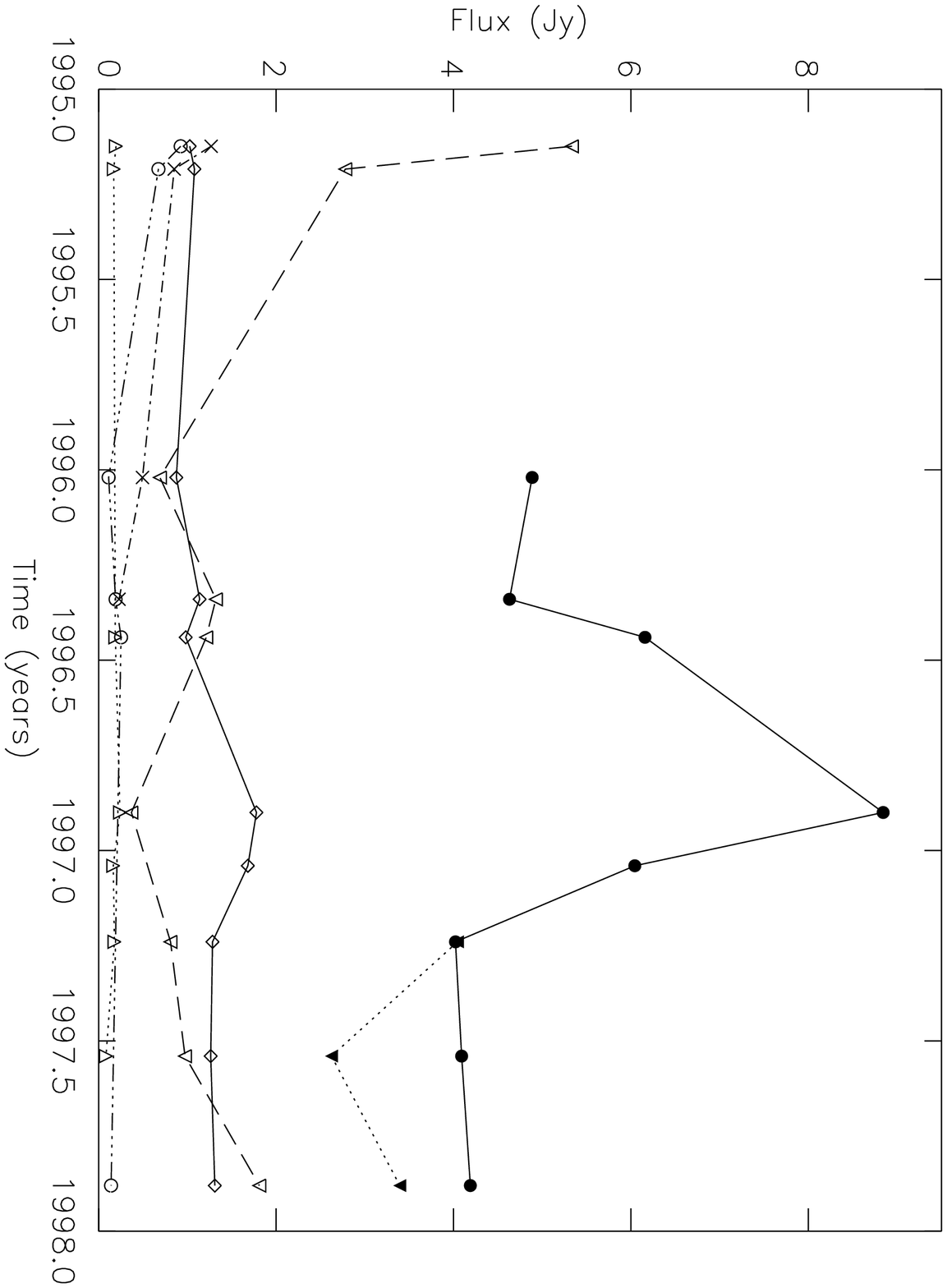}{3.375 in}{90}{60}{60}{227}{-27}
\begin{center}
FIG. 10.---{\em Continued}
\end{center}
\end{figure*}

The VLBI data are clearly undersampled, even though 3C\,279 is one of the
best-observed AGNs with VLBI.  However, the obvious correlation
between the VLBI core and total flux light curves shows that flux variations in the core are largely
responsible for the total flux variability.  
The logarithmic contours on the image series (Figures 1 to 4) tend to
suppress these changes, with the result that the image evolution appears to
be largely uncorrelated.  A consequence of this is that relating flux
variations in the more distant components to the total flux will be very
difficult, because they are masked by the much stronger core variations.
This behavior of 3C\,279 (where the total flux variability is dominated by the
VLBI core and the jet components are much weaker than the core) is quite different
from its behavior in the early 1980s, when the total flux density variations
were dominated by component C3 (U89), and
from that of 3C\,345, where the total flux variability has
been linked to new jet components which have all surpassed the
core in brightness (Valtaoja et al. 1999).

The flux history of the VLBI components generally shows a fading with time,
with component lifetimes of roughly 3 years.  Exceptions to this are
components C4, which has maintained an approximately constant flux over the 
six year monitoring period and shows no signs of fading, and C5, which at first
faded but then maintained an approximately constant flux.
Components C6 and C7a both fade at first but then show secondary increases
in flux; such an increase could be due to a changing Doppler factor along a slightly curving path.
3C\,279 ejected an exceptionally
bright component (C8) in late 1995.  This component was first detected with a flux
density of 5 Jy, it then rose in flux until it reached 9 Jy (making it the single
exception to our previous statement that the jet components in 3C\,279 were
generally much weaker than the core),
it later faded to a flux density of 4 Jy at the end of our monitoring
in 1997.

\subsection{Radio Spectrum}
The 22 and 43 GHz images and model fits can be used to calculate two-point spectral
indices.  Considering first the model-fit Gaussian fluxes, we calculate mean
values for the spectral index of the core and C4 to be $-0.16$ and $-0.70$
respectively, for $S \propto \nu^{+\alpha}$.
Variations in spectral index from epoch to epoch are difficult to distinguish from flux errors
due to calibration or model fitting, making it difficult to draw conclusions about spectral
index variations with time from these observations.  Similarly, it is difficult to 
quantify the spectral index of the inner jet components from the model fits, since the
inner jet is a complex area with multiple components and subject to component blending
at 22 GHz.

The approximately flat index for the core indicates a synchrotron turnover frequency between 22 and 43 GHz.
This is consistent with the results of Grandi et al. (1996), who concluded the turnover
frequency lay between 37 and 90 GHz based on total flux density monitoring from 1992 December to 1993 January.
The turnover frequency of the core is variable; U89 found
a turnover frequency around 11 GHz based on their three-frequency VLBI observations from 1983.
The steep spectral index of C4 indicates that this component is optically thin above 22 GHz.
Piner et al. (2000a) found a spectral index for C4 of $+0.25$ between 1.6 and 5 GHz, placing
the turnover frequency for C4 between 5 and 22 GHz.    

\subsection{Radio Core Brightness Temperatures}
\label{tbsec}
\subsubsection{Brightness Temperature Estimates}
\label{tbmeassec}
Brightness temperature values can be used to constrain
both Doppler beaming factors and physical processes occurring in a source.
The maximum brightness temperature of a circular Gaussian\footnote{
A Gaussian brightness distribution is the only brightness
distribution fully supported in DIFMAP.} is given by
\begin{equation}
\label{tbeq}
T_{B}=1.22\times10^{12}\;\frac{S(1+z)}{a^{2}\nu^{2}}\;\rm{K},
\end{equation}
where $S$ is the flux density of the Gaussian in Janskys,
$a$ is the FWHM of the Gaussian in mas,
$\nu$ is the observation frequency in GHz, and $z$ is the redshift.
A fact that is often neglected is that the minimum measurable size depends on
the signal-to-noise of the observations, because a smaller slope in the amplitude vs.
$(u,v)$ distance plane can be measured when the error bars on the visibilities are smaller.
An expression for the minimum measurable size 
in the visibility plane is given by Lovell et al. (2000) to be
\begin{equation}
\label{sizeeq}
a<\frac{2.4\sqrt{\frac{N}{S}}}{U}\;\rm{mas},
\end{equation}
where $N$ is the integrated rms noise of the observations in Janskys and $U$ is the maximum
baseline length in units of 100 M$\lambda$.  
Expressed in terms of a uniformly weighted beam size, $\theta_{beam}\approx 1.34\times10^{8}\lambda/d$ mas,
this is $a<1.8\sqrt{\frac{N}{S}}\theta_{beam}$ mas.  Inserting equation~(\ref{sizeeq}) into
equation~(\ref{tbeq}), the maximum measurable brightness temperature is
\begin{equation}
\label{newtbeq}
T_{B}=235\;\frac{S^{2}D^{2}(1+z)}{N}\;\rm{K},
\end{equation}
where $D$ is the maximum baseline length in km.  The maximum measurable brightness temperature
depends on the source flux squared, because of the dependence of minimum measurable size
on the SNR.  For an observation of 3C\,279 with the VLBA at 22 GHz with
$S\sim10$ Jy, $N\sim10$ mJy, and $D\approx8600$ km, the minimum measurable size is about $\frac{1}{20}$
of a beam, and the maximum measurable brightness temperature is over $10^{14}$ K.  The commonly
quoted limit of $\sim10^{12}$ K for earth-size baselines is based on a canonical 1 Jy source and
a minimum measurable size of about a third of a beam.

Brightness temperatures for the core of 3C\,279 from our 22 and 43 GHz
model fits are listed in Table~\ref{tbtab}.  Upper and lower limits to the brightness
temperature at each epoch are also given; these upper and lower limits were calculated
using the Difwrap program (Lovell 2000) as described by Piner et al. (2000a).
In some cases, a model-fit brightness temperature in excess of that
given by equation~(\ref{newtbeq}) is obtained, these values are indistinguishable
from an infinite brightness temperature, i.e., the component cannot be resolved
with the given data.  In these cases, the lower limit to the brightness temperature
is a more meaningful number than the best-fit value.

\begin{table*}[!t]
\caption{3C\,279 Radio Core Gaussian Brightness Temperatures}
\label{tbtab}
\begin{center}
{\small \begin{tabular}{l r r r | r r r} \tableline \tableline
& \multicolumn{3}{c}{22 GHz} & \multicolumn{3}{c}{43 GHz} \\ \tableline 
& & \multicolumn{1}{c}{Min.\tablenotemark{a}} & \multicolumn{1}{c}{Max.\tablenotemark{b}} & 
& \multicolumn{1}{c}{Min.\tablenotemark{a}} & \multicolumn{1}{c}{Max.\tablenotemark{b}} \\ 
& \multicolumn{1}{c}{$T_{B}$} & \multicolumn{1}{c}{$T_{B}$} & \multicolumn{1}{c}{$T_{B}$}
& \multicolumn{1}{c}{$T_{B}$} & \multicolumn{1}{c}{$T_{B}$} & \multicolumn{1}{c}{$T_{B}$} \\ 
Epoch & \multicolumn{1}{c}{($10^{12}$ K)} & \multicolumn{1}{c}{($10^{12}$ K)} & \multicolumn{1}{c}{($10^{12}$ K)}
& \multicolumn{1}{c}{($10^{12}$ K)} & \multicolumn{1}{c}{($10^{12}$ K)} & \multicolumn{1}{c}{($10^{12}$ K)} \\ \tableline
1991 Jun 24 &  3.8     &  2.3 & 20.8     &  ...     & ... &  ...     \\ 
1992 Jun 14 &  3.4     &  2.0 &  8.1     &  ...     & ... &  ...     \\ 
1992 Nov 10 & 51.5     &  5.4 & $\infty$ &  ...     & ... &  ...     \\ 
1993 Feb 17 &  5.5     &  3.0 & 11.5     &  ...     & ... &  ...     \\ 
1993 Nov 8  & 11.6     &  6.3 & $\infty$ &  ...     & ... &  ...     \\ 
1994 Mar 2  &  6.8     &  4.6 & 10.6     &  ...     & ... &  ...     \\ 
1994 Jun 12 &  6.4     &  3.8 & 12.5     &  ...     & ... &  ...     \\ 
1994 Sep 21 &  5.6     &  3.5 &  8.0     &  ...     & ... &  ...     \\ 
1995 Jan 4  &  9.1     &  4.8 & 15.2     &  ...     & ... &  ...     \\ 
1995 Feb 25 &  6.3     &  3.0 & 18.4     & $\infty$ & 4.7 & $\infty$ \\ 
1995 Mar 19 &  7.1     &  4.9 & 11.0     & 12.6     & 7.6 & 20.0     \\ 
1996 Jan 7  & 13.3     &  8.9 & 27.1     &  7.5     & 5.0 &  9.4     \\ 
1996 May 4  &  ...     &  ... &  ...     & 19.0     & 6.6 & $\infty$ \\ 
1996 May 13 & 15.5     & 12.1 & 25.2     &  ...     & ... &  ...     \\ 
1996 Jun 9  & 38.2     & 18.4 & 116      &  7.4     & 5.6 &  8.9     \\ 
1996 Nov 24 &  ...     &  ... &  ...     &  5.4     & 3.9 & 11.0     \\ 
1997 Jan 15 & $\infty$ & 11.0 & $\infty$ & 18.1     & 7.0 & $\infty$ \\ 
1997 Mar 29 & $\infty$ & 11.7 & $\infty$ &  2.9     & 2.2 &  3.9     \\ 
1997 Jul 16 & 21.9     &  8.0 & $\infty$ &  4.1     & 2.5 &  5.7     \\ 
1997 Nov 16 & 13.4     &  7.4 & 21.3     &  9.1     & 6.7 & 13.0     \\ \tableline
\end{tabular}}
\end{center}
\tablenotetext{a}{Lower limit to the brightness temperature calculated using the Difwrap program.}
\tablenotetext{b}{Upper limit to the brightness temperature calculated using the Difwrap program.
In all cases where a numerical value is quoted rather than infinity, this value is less
than the maximum measurable brightness temperature given by equation~(\ref{newtbeq}).}
\end{table*}

Note that allowable ranges for the core size and flux were derived
for visibilities self-calibrated to the CLEAN images shown in this paper.
We performed several tests to assess how self-calibrating to the
variable Gaussian models could affect the allowed ranges of core size and core flux.
Repeated self-calibration and model fitting to the Gaussian models
from Table~\ref{mfittab} with the core size or flux displaced from
the best-fit value allowed larger ranges of these parameters to fit the visibilities,
but Gaussian models obtained in this fashion no longer accurately represented the corresponding CLEAN image.
When only a single Gaussian at the core location was used --- with the
rest of the source structure represented by fixed CLEAN components to retain the constraints
provided by the hybrid mapping procedure --- then repeated self-calibration and model fitting to this Gaussian,
with the size or flux displaced from the best-fit value, produced about the same range
of allowable brightness temperatures as those given in Table~\ref{tbtab}.

The brightness temperatures at 22 GHz in Table~\ref{tbtab} can be divided into two time ranges:
from 1991 through 1995 the upper and lower limits are consistent
with a constant brightness temperature between 6.3 and $8.0\times10^{12}$~K;
after 1995 the upper and lower limits are consistent
with a constant brightness temperature between 18.4 and $21.3\times10^{12}$~K,
indicating an increase in the brightness temperature sometime during 1995.
At 43 GHz all brightness temperature limits are consistent with 
a constant brightness temperature between 7.6 and $8.9\times10^{12}$~K, with the exception
of the 1997 Mar 29 and 1997 Jul 16 epochs which have lower brightness temperatures.
For comparison, Piner et al. (2000a) found a maximum brightness temperature 
lower limit of $5\times10^{12}$~K
for 3C\,279 from VSOP observations at 1.6 GHz.  For observations at the same frequencies, VSOP should be
able to measure brightness temperatures about a factor of two higher than the VLBA (from inserting
relevant baseline lengths and sensitivities into equation~[\ref{newtbeq}]); however, the inverted spectrum
of 3C\,279 means that a higher brightness temperature can be measured with the VLBA near the turnover
frequency than at lower frequencies with VSOP.

The core of 3C\,279 is partially resolved in the majority of these observations.  The high brightness temperatures estimated
at 22 GHz of $\sim2\times10^{13}$~K are considerably less than the maximum measurable brightness temperature
derived above for these observations of over $10^{14}$ K.  The estimated sizes are also larger than the minimum
measurable sizes given by equation~(\ref{sizeeq}), and a considerable worsening of the fit is found at
many epochs for a zero-size Gaussian.  Average core sizes are 0.07 mas at 22 GHz and 0.045 mas
at 43 GHz.  These sizes are consistent with an inhomogeneous jet model such as the K\"{o}nigl model
(K\"{o}nigl 1981), where the ``core'' at different frequencies is viewed further down an expanding jet,
with the minimum size occurring at the turnover frequency.

\subsubsection{Brightness Temperature Implications}
\label{tbimpsec}
The estimated brightness temperature of $\sim2\times10^{13}$~K after 1995 at 22 GHz is among the
highest direct estimates of brightness temperature (using a size rather than a variability
timescale).  Similar high brightness temperatures of a few times $10^{12}$ to
$10^{13}$ K have been recorded in observations of other sources by the VSOP mission (Bower \& Backer 1998;
Lovell et al. 2000; Preston et al. 2000; Shen et al. 1999).
Observed brightness temperatures are increased by the Doppler factor, and there are
several mechanisms that may act to limit the rest-frame brightness temperature of a synchrotron source:
rapid energy loss by inverse Compton emission that limits the
brightness temperature to $\sim5\times10^{11} - 1\times10^{12}$~K (Kellermann \& Pauliny-Toth 1969), induced Compton
scattering that limits the brightness temperature to $\sim2\times10^{11}$~K (Sincell \& Krolik 1994), and
equipartition of energy between particles and fields that limits the brightness temperature to
$\sim5\times10^{10} - 1\times10^{11}$~K (Readhead 1994).
Since these limiting brightness temperatures were all derived for homogeneous optically thick
spheres, our Gaussian brightness temperatures should be converted to homogeneous sphere brightness
temperatures for comparison, by multiplying by the appropriate conversion factor of 0.56,
which is valid for partially resolved components 
(Pearson 1995; Hirabayashi et al. 1998, in particular the correction in the erratum to this paper).
The highest brightness temperatures at 22 GHz are then $\sim1\times10^{13}$~K, implying 
Doppler factors of 10-20, 50, or 140 respectively for the three limiting cases discussed above
(the actual equipartition brightness temperature for 3C\,279 from equation (4a) of Readhead [1994]
is $\sim7\times10^{10}$~K).    

Monte Carlo simulations of the EGRET blazar sample by Lister (1998) predict a Doppler factor
distribution with a tail extending up to $\delta\approx50$.  However, these simulations were based on
a maximum observed jet speed of 15$h^{-1}c$.  Since recent results on the apparent speeds in EGRET
blazars (Marscher et al. 2000) show speeds up to $\approx30h^{-1}c$ ($\approx45c$ for $h=0.65$), revised simulations 
should show Doppler factors extending up to $\delta\approx100$.  
Such high Doppler factor sources would be intrinsically rare as evidenced by the absence
of such high speeds in radio-selected samples (Kellermann et al. 2000), and would only be present in
samples that have a stronger orientation bias as has been suggested for the EGRET sample (Marscher et al. 2000).
When combined with the apparent jet speeds in 3C\,279 ($\S$~\ref{radmotion}), a Doppler factor $\delta\sim100$ 
implies an extremely small viewing angle and an extremely small opening angle.
If the Doppler factor is this high in the radio core then it is likely reduced by a 
slight bend of the radio jet away from the line-of-sight by the time VLBI scales have been reached.  Note that the
Doppler factor shown for C4 in Figure 8$e$ ($\S$~\ref{3dc4}) is not a measurement, since we assumed
a constant $\Gamma$ close to the minimum value allowed by the observed speed of C4.  A higher
$\Gamma$ raises $\delta$ in Figure 8$e$, decreases viewing angle in Figure 8$d$, and stretches the
vertical scale in Figure 8$f$.

If an independent measurement of another function of the Doppler factor exists
then the rest-frame brightness temperature and Doppler factor can be found.  Such a quantity is the variability
brightness temperature (L\"{a}hteenm\"{a}ki, Valtaoja, \& Wiik 1999).  The variability brightness
temperature measured for 3C\,279 is $2.3\times10^{14}$~K (L\"{a}hteenm\"{a}ki \& Valtaoja 1999), and
was measured at 37 GHz for a flare peaking at 1995.2
(L\"{a}hteenm\"{a}ki 1999, private communication).  
Comparing this to the lower limit to the 43~GHz VLBI brightness temperature
at this epoch ($4.3\times10^{12}$~K for a homogeneous sphere), 
we obtain a rest-frame brightness temperature of $6\times10^{11}$~K
and a Doppler factor of 7.4 using equations from L\"{a}hteenm\"{a}ki et al. (1999).  
This implies that the jet of 3C\,279 is particle-dominated by many orders of
magnitude, and the brightness temperature is limited
by inverse Compton effects.  Slysh (1992) argues that such a high rest-frame brightness temperature
can be maintained on a timescale of years following the injection of electrons of sufficiently high energy,
or can be maintained indefinitely with continuous in situ electron acceleration.
In this case, the Doppler factor for the 3C\,279 core is at the
low end of the distribution expected for EGRET sources (Lister 1998), and the jet must bend toward the 
line-of-sight as it moves from the core to C4, since the Doppler factor shown for C4 in Figure 8$e$
is a lower limit for this component.

The Doppler factors and rest-frame brightness temperatures suggested by equipartition and by the variability
brightness temperature are discrepant for 3C\,279.  This suggests that one of the following situations
is true: 
\begin{enumerate}
\item{3C\,279 is approximately in equipartition and has an extremely high Doppler factor
($\delta\sim100$) in the tail end of the distribution expected for EGRET blazars
from the simulations of Lister (1998).  This means
the variability brightness temperature measured for 3C\,279 is a lower limit and the actual
variability brightness temperature is much larger (calculated rest-frame brightness temperature goes down and
$\delta$ goes up as the variability brightness temperature increases for a given VLBI brightness temperature,
see equations (6)-(8) of L\"{a}hteenmaki et al. [1999]).
This could only occur if extremely rapid variations in the 3C\,279 light curve had been missed.}
\item{The measured variability brightness temperature is correct and 3C\,279 has a strongly 
particle-dominated parsec-scale jet where the rest-frame brightness temperature is high and limited by 
inverse Compton effects.  In this case the Doppler factor for the 3C\,279 core would be at the
low end of the distribution expected for EGRET sources from the simulations of Lister (1998).}
\end{enumerate}

\section{CONCLUSIONS}
\label{conclusions}

We have presented the results from our intensive high-frequency VLBI monitoring
of 3C\,279 during the years 1991 to 1997.  Three major results of this study are:
\begin{description}
\item[1.]{Apparent speeds measured for six superluminal components range from
4.8 to 7.5$c$.}
\item[2.]{Comparison of VLBI and single-dish light curves 
from Mets\"{a}hovi show that variations in the total flux density light curves
can mainly be accounted for by changes in the VLBI core flux density, not the jet. 
This contrasts both with 3C\,279's behavior in the early 1980s (U89)
and with that of 3C\,345 (Valtaoja et al. 1999), because in both of these cases total flux density increases
were linked to the youngest jet component.}
\item[3.]{
The VLBI core is partially resolved and has a size of $\sim$0.07 mas at 22 GHz and $\sim$0.045 mas at 43 GHz.
The uniform-sphere brightness temperature at 22 GHz is $\sim1\times10^{13}$~K after 1995,
which is one of the highest direct estimates of a brightness temperature.
If the variability brightness temperature measured for 3C\,279 by L\"{a}hteenm\"{a}ki \& Valtaoja (1999)
is an actual value and not a lower limit, then the rest-frame brightness temperature of 3C\,279 is
quite high and limited by inverse Compton effects rather than equipartition.}
\end{description}
Other conclusions are:
\begin{description}
\item[4.]{Component C4 has been very long-lived and was detected at all epochs.
C4 followed a curved path,
and we were able to reconstruct its 3-dimensional trajectory by fitting polynomials to
its position versus time.  The reconstructed trajectory is suggestive of low pitch
angle helical motion, and constrains the path of C4 to lie within $\sim2\arcdeg$ of the line-of-sight.}
\item[5.]{There is a stationary component at $\sim$1 mas that fades and may move slightly inward
during the course of our observations.  This behavior could be due to a previous interaction
with C4, and resembles the interactions between superluminal and stationary components seen in
simulations by G\'{o}mez et al. (1997).}
\item[6.]{Components in the inner jet were relatively short-lived, and faded by the time they reached
$\sim$1 mas from the core.}  
\item[7.]{The VLBI components have different speeds and
position angles from each other, but these differences do not match the differences
predicted by the precession model of Abraham \& Carrara (1998).  The position angles also differ
from that of the larger-scale structure. The longer-lived component C4 shows no signs of altering its position angle
to match that of the larger-scale structure.}
\item[8.]{The intrinsic jet opening angle is constrained to be smaller than about half a degree
by the size and reconstructed trajectory of C4.}
\item[9.]{
Although VLBI components
were born about six months prior to each of the two observed $\gamma$-ray high states,
the sparseness of the $\gamma$-ray data prevents a statistical analysis of
possible correlations.}
\item[10.]{The VLBI core spectrum indicates the core turnover frequency is between 22 and 43
GHz.  Comparison of C4's flux to that at lower frequencies suggests a turnover frequency
for this component between 5 and 22 GHz.}
\end{description}

Completion of this large number of epochs moves 3C\,279 into
the category of sources (such as 3C\,345) for which large VLBI databases exist for detailed studies
of jet properties.  We suggest readers view the animations made from these observations at
ftp://sgra.jpl.nasa.gov/pub/users/glenn/3c279\_22ghz.mpeg
and 3c279\_43ghz.mpeg to fully appreciate the complexity of this source
as revealed by multi-epoch monitoring.
Possible studies using this large amount of data could include  
application of specific
hydrodynamic instability or magneto-hydrodynamic models to the
motion of C4, or application of precession models to the differing
speeds and position angles of components using much more data
than was used for the precession model of Abraham \& Carrara (1998).
Future data reduction will include
performing the polarization analysis of the polarization-sensitive
observations listed in Table~\ref{vlbiobs} and analyzing all unreduced lower-frequency
VLBI observations.  Together with data from this paper, X-ray data, and
$\gamma$-ray data, this should allow us to test inverse Compton models against the multiwavelength
spectrum (Piner et al. 2000b).  It will be important to continue monitoring 3C\,279
with the VLBA throughout the years between the EGRET and GLAST missions, so that
a deeper understanding of this canonical $\gamma$-ray blazar
can be obtained before the launch of the next $\gamma$-ray mission.

\acknowledgments

Part of the work described in this paper has been carried out at the Jet
Propulsion Laboratory, California Institute of Technology, under
contract with the National Aeronautics and Space Administration.
This research has made use of data from the University of Michigan Radio
Astronomy Observatory which is supported by the National Science Foundation
(grant AST-9421979) and by funds from the University of Michigan.  We are
grateful to Bob Hartman, Ken Kellermann, and Anne L\"{a}hteenm\"{a}ki for
providing data by private communication.
The National Radio Astronomy Observatory is a facility of the
National Science Foundation, operated under cooperative agreement by
Associated Universities, Inc. This work was supported in part by NSF grant
AST\,9117100, NASA grants NAG\,5-2167 and 7-1260, and the NASA Long Term
Space Astrophysics program.

\begin{center}
{\bf APPENDIX A\\}
{\bf 22 GHZ VLBI OBSERVATIONS FROM THE 1980S}
\end{center}
In this appendix we present early 22 GHz VLBI observations from our monitoring program.
These data have been previously published by U89 and C93,
but we have reimaged the data using the interactive imaging and editing capabilities of DIFMAP,
which has resulted in significant improvements.  These early VLBI observations are listed
in Table A1, which is analogous to Table~\ref{vlbiobs} in the main text.  
The remade images are shown in Figure A1; these
should now be used in place of the corresponding images from U89 and C93.
The parameters of these images are listed in Table A2, which is analogous
to Table~\ref{imtab} in the main text.
The low SNR of these observations and the high ellipticity of the beams causes model fitting
to produce ambiguous results.  This, together with the three-year gap between the 1988 and 1991 epochs,
means that results from these images cannot be consistently connected to the image series presented
in the main text.

\begin{table*}[!t]
\begin{center}
{\footnotesize TABLE A1 \\ 
1980S 22 GHZ VLBI OBSERVATIONS \\ }
{\scriptsize \begin{tabular}{l l l l c c c c} \tableline \tableline
& \multicolumn{1}{c}{Experiment} & & & \multicolumn{1}{c}{Bandwidth} &
Obs. Time & Frequencies & \\
\multicolumn{1}{c}{Epoch} & \multicolumn{1}{c}{Name} & VLBA Antennas
& Other antennas\tablenotemark{a} & \multicolumn{1}{c}{(MHz)} &
(minutes) & (GHz) & Polarization \\ \tableline
1984 Oct 2  & ...     & ... & Eb,Gb,Hs,Mp,On,Ov,Y1 & 2 & 454 & 22 & LCP \\ 
1985 Oct 1  & U15G    & ... & Eb,Gb,Hs,Mp,On,Ov,Y1 & 2 & 384 & 22 & LCP \\ 
1987 Jun 4  & Z13G-AH & ... & Eb,Gb,Hs,Mc,On,Ov,Y1 & 2 & 569 & 22 & LCP \\  
1988 Feb 28 & U17G    & ... & Eb,Gb,Hs,On,Ov       & 2 & 616 & 22 & LCP \\ \tableline
\end{tabular}}
\end{center}
\tablenotetext{a}{Antenna locations and sizes are as follows:
Eb = Effelsberg, Germany, 100 m;
Gb = Green Bank, WV, 43 m; Hs = Haystack, MA, 37 m;
Mc = Medicina, Italy, 32 m;
Mp = Maryland Point, MD, 26 m;
On = Onsala, Sweden, 20 m;
Ov = Owens Valley, CA, 40m;
Y1 = one antenna of the VLA, Socorro, NM, 25 m.}
\end{table*}

\begin{table*}
\begin{center}
{\footnotesize TABLE A2 \\ 
1980S 22 GHZ IMAGE PARAMETERS \\ }
{\scriptsize \begin{tabular}{l c l c c c c c} \tableline \tableline
& & & Total & CLEAN & Peak & Lowest & Contours\tablenotemark{d} \\ 
& Frequency & & Flux\tablenotemark{b} & Flux & Flux & Contour\tablenotemark{c} & (multiples of \\ 
\multicolumn{1}{c}{Epoch} & (GHz) & \multicolumn{1}{c}{Beam\tablenotemark{a}} & (Jy) & (Jy) &
(Jy beam$^{-1}$) & (mJy beam$^{-1}$) & lowest contour) \\ \tableline
1984 Oct 2  & 22 & 3.35,0.27,$-$8.1  &  9.7 &  9.8 & 5.8  & 103   & 1...2$^{5}$  \\ 
1985 Oct 1  & 22 & 3.26,0.24,$-$8.9  &  9.2 &  9.1 & 5.1  & 71.9  & 1...2$^{6}$  \\ 
1987 Jun 4  & 22 & 1.54,0.20,$-$10.3 & 10.4 & 10.1 & 6.0  & 149   & 1...2$^{5}$  \\ 
1988 Feb 28 & 22 & 3.47,0.22,$-$8.1  & 11.9 & 11.2 & 6.5  & 48.6  & 1...2$^{7}$  \\ \tableline
\end{tabular}}
\end{center}
\tablenotetext{a}{Numbers given for the beam are the FWHMs of the major
and minor axes in mas, and the position angle of the major axis in degrees.
The beam has been synthesized using uniform weighting.}
\tablenotetext{b}{Single-dish flux from Mets\"{a}hovi at 22 or 37 GHz.}
\tablenotetext{c}{The lowest contour is set to be three times the rms noise
in the full image.}
\tablenotetext{d}{Contour levels are represented by the geometric series 1...2$^{n}$,
e.g. for $n=5$ the contour levels would be $\pm$1,2,4,8,16,32.}
\end{table*}

\begin{figure*}
\plotfiddle{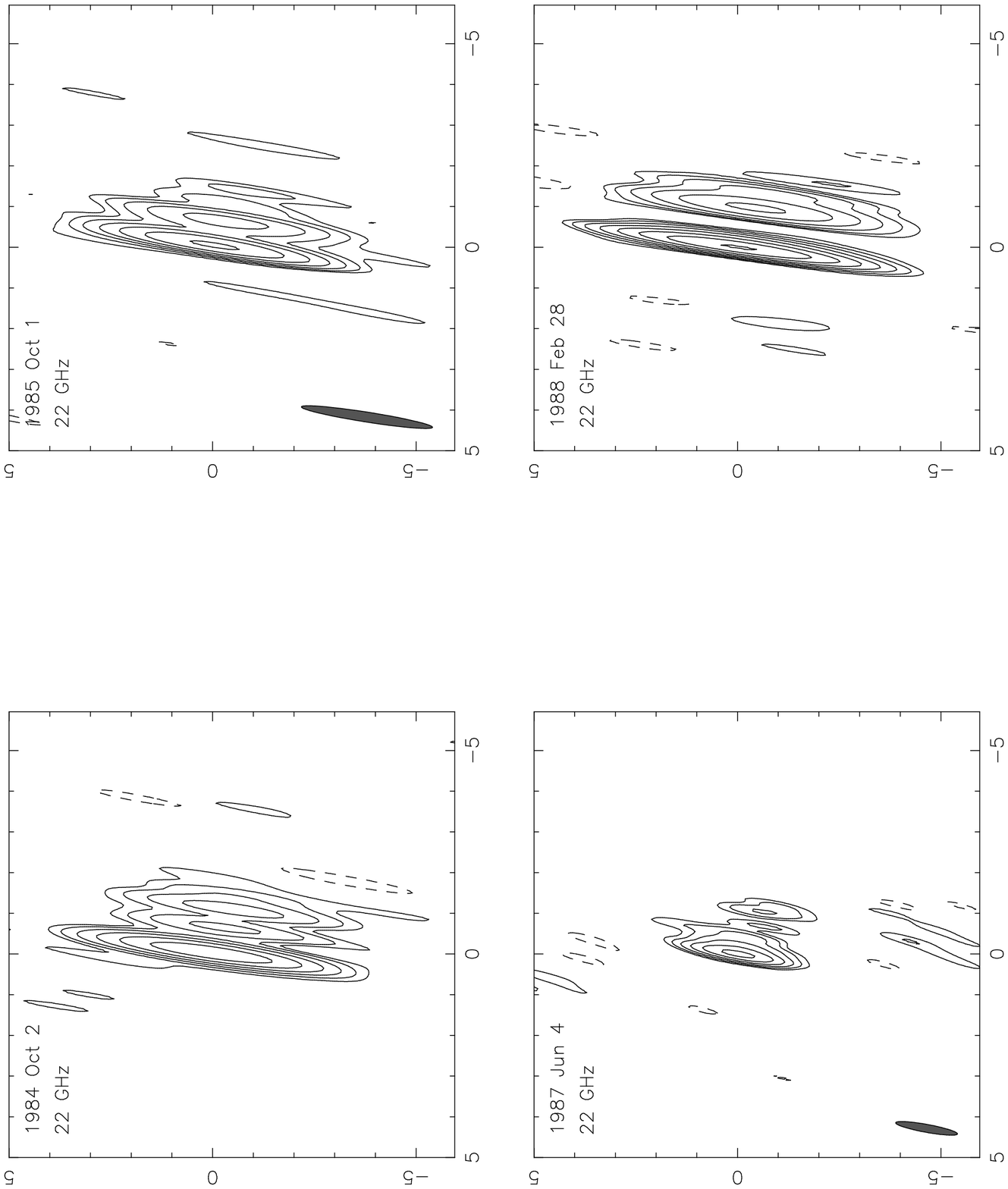}{4.25in}{-90}{55}{55}{41}{336}
FIG. A1.---22 GHz uniformly weighted images of 3C\,279 from the 4 epochs listed in Table~A1.
The axes are labeled in milliarcseconds.
Parameters of the images are given in Table~A2.
\end{figure*}

\end{document}